\documentclass{aa}  

\usepackage{graphicx}
\usepackage{amsmath}    % Advanced maths commands
\usepackage{amssymb}    % Extra maths symbols
\usepackage{txfonts}
\usepackage{natbib,twoopt}
\usepackage[breaklinks=true]{hyperref} %% to avoid \citeads line fills
\usepackage{float}
\usepackage[caption = false]{subfig}

\hypersetup{
  colorlinks=true,   %% links colored instead of frames
  urlcolor=blue,     %% external hyperlinks
  linkcolor=blue,     %% internal latex links (eg Fig)
}

\bibpunct{(}{)}{;}{a}{}{,} %% natbib format for A&A and ApJ

\makeatletter
\newcommandtwoopt{\citeads}[3][][]{\href{http://adsabs.harvard.edu/abs/#3}%
{\def\hyper@linkstart##1##2{}%
\let\hyper@linkend\@empty\citealp[#1][#2]{#3}}}
\newcommandtwoopt{\citepads}[3][][]{\href{http://adsabs.harvard.edu/abs/#3}%
{\def\hyper@linkstart##1##2{}%
\let\hyper@linkend\@empty\citep[#1][#2]{#3}}}
\newcommandtwoopt{\citetads}[3][][]{\href{http://adsabs.harvard.edu/abs/#3}%
{\def\hyper@linkstart##1##2{}%
\let\hyper@linkend\@empty\citet[#1][#2]{#3}}}
\newcommandtwoopt{\citeyearads}[3][][]%
{\href{http://adsabs.harvard.edu/abs/#3}
{\def\hyper@linkstart##1##2{}%
\let\hyper@linkend\@empty\citeyear[#1][#2]{#3}}}
\makeatother

\newcommand{\lo}{\log L_{\rm UV}}
\newcommand{\lx}{\log L_{\rm X}}
\newcommand{\fo}{\log F_{\rm UV}}
\newcommand{\fx}{\log F_{\rm X}}
\newcommand{\Lo}{L_{\rm UV}}
\newcommand{\Fo}{F_{\rm UV}}
\newcommand{\Lx}{L_{\rm X}}
\newcommand{\Fx}{F_{\rm X}}
\newcommand{\nuo}{\nu_{\rm o}}
\newcommand{\nux}{\nu_{\rm X}}
\newcommand{\aox}{\alpha_{\rm ox}}
\newcommand{\mbh}{M_{\rm BH}}
\newcommand{\lbol}{L_{\rm bol}}

\newcommand{\gammax}{\Gamma_{\rm X}}
\newcommand{\sigmaB}{\sigma_{\rm B}}
\newcommand{\sigmaT}{\sigma_{\rm T}}
\newcommand{\kB}{k_{\rm B}}
\newcommand{\Rs}{R_{\rm S}}
\newcommand{\rg}{r_{\rm g}}
\newcommand{\fmin}{F_{\rm min}}

\newcommand{\w}{\upsilon_{\rm fwhm}}
\newcommand{\wtwo}{\upsilon_{\rm{fwhm},2000}}

\newcommand{\rev}[1]{{ #1}}
\newcommand{\revs}[1]{{ #1}}
\newcommand{\revt}[1]{{ #1}}
\newcommand{\revle}[1]{{ #1}}

\begin{document}

%%%%% AUTHORS - PLACE YOUR OWN PACKAGES HERE %%%%%

%%%%%%%%%%%%%%%%%%% TITLE PAGE %%%%%%%%%%%%%%%%%%%

% Title of the paper, and the short title which is used in the headers.
% Keep the title short and informative.
\title{Quasars as standard candles I: The physical relation between disc and coronal emission}

% The list of authors, and the short list which is used in the headers.
% If you need two or more lines of authors, add an extra line using \newauthor
\author{
E. Lusso\inst{1,2}\thanks{E-mail: \revle{elisabeta.lusso@durham.ac.uk}} and G. Risaliti\inst{3,2}}
\institute{
% List of institutions
$^{1}$Centre for Extragalactic Astronomy, Durham University, South Road, Durham, DH1 3LE, UK \\
$^{2}$INAF--Osservatorio Astrofisico di Arcetri, 50125 Florence, Italy\\
$^{3}$Dipartimento di Fisica e Astronomia, Universit\'a di Firenze, via G. Sansone 1, 50019 Sesto Fiorentino, Firenze, Italy
}

% These dates will be filled out by the publisher
\date{Accepted 10 March 2017/Received 17 November 2016}

%-------------------------------------------------------------------

\abstract
   {A tight non-linear relation exists between the X-ray and UV emission in quasars (i.e. $\Lx\propto\Lo^{\gamma}$), with a dispersion of $\sim$0.2~dex over\rev{approximately three~orders of magnitude in luminosity}. Such observational evidence has two relevant consequences: (1) an ubiquitous physical mechanism must regulate the energy transfer from the accretion disc to the X-ray emitting {\it corona}, and (2) the non-linearity of the relation provides a new, powerful way to estimate the absolute luminosity, turning quasars into a new class of {\it standard candles}.}
  {Here we propose a modified version of this relation, which involves the emission line full-width half maximum, $\Lx\propto\Lo^{\hat\gamma}\w^{\hat\beta}$.}
   {We interpret this new relation through a simple, {\it ad-hoc} model of accretion disc corona, derived from previous works in the literature where it is assumed that reconnection and magnetic loops above the accretion disc can account for the production of the primary X-ray radiation.}
   {We find that the monochromatic optical-UV (2500 \AA) and X--ray (2 keV) luminosities depend on the black hole mass and accretion rate as $\Lo\propto \mbh^{4/3} (\dot{M}/\dot{M}_{\rm Edd})^{2/3}$ and $\Lx\propto \mbh^{19/21} (\dot{M}/\dot{M}_{\rm Edd})^{5/21}$, respectively.
Assuming a broad line region size function of the disc luminosity $R_{\rm blr}\propto L_{\rm disc}^{0.5}$ we finally have that $\Lx\propto\Lo^{4/7} \w^{4/7}$. Such relation is remarkably consistent with the slopes and the normalisation obtained from a fit of a sample of 545 optically selected quasars from SDSS DR7 cross matched with the latest XMM--{\it Newton} catalogue 3XMM-DR6.}
  {\rev{The homogeneous sample used here has a dispersion of 0.21 dex, which is much lower than previous works in the literature and suggests a tight physical relation between the accretion disc and the X-ray emitting corona. We also obtained a possible physical interpretation of the $\Lx-\Lo$ relation (considering also the effect of $\w$), which puts the determination of distances based on this relation on a sounder physical grounds. The proposed new relation does not evolve with time, and thus it can be employed as a cosmological indicator to robustly estimate cosmological parameters.}}


   \maketitle
%
%-------------------------------------------------------------------

\section{Introduction}
%Accreting super massive black holes (SMBHs) in the centre of galaxies represent only a small fraction ($\sim$10\%) of the total galaxy population, yet these active galactic nuclei (AGN) denote a specific and important stage of galaxy evolution. AGN are observed to correlate with the galaxy properties such as the mass of the bulge of the host-galaxy (\citeads{1998AJ....115.2285M,2003ApJ...589L..21M}), with the velocity dispersion of the bulge (\citeads{2000ApJ...539L...9F,2002ApJ...574..740T}), and with the luminosity of the bulge (\citeads{1995ARA&A..33..581K}).

% general sed introduction
Accreting super massive black holes (SMBHs) in the centre of galaxies (i.e. {active galactic nuclei}  (AGN)) display characteristic observational features over a wide range of frequencies. Specifically, the AGN spectral energy distribution (SED) shows significant emission in the optical-UV, the so-called big blue bump (BBB), with a softening at higher energies \citepads{1989ApJ...347...29S,1994ApJS...95....1E,2006ApJS..166..470R,2007AJ....133.1780T,2011ApJS..196....2S,2012ApJ...759....6E}. This emission is thought to origine from an optically thick disc surrounding the SMBH \citepads{1973A&A....24..337S,1973blho.conf..343N}. The disc temperature goes from $\sim5\times10^5$K to $\sim8.7\times10^4$K for a SMBH of $10^6$ and $10^9M_\odot$ (at a given accretion) respectively, which corresponds to a peak emission in the range $\log \nu/{\rm Hz}\simeq 15-16.5$\footnote{A basic prediction of simple accretion disc models is that the disc temperature decreases as the black hole mass increases \citepads{1973A&A....24..337S}.}.

At X-ray energies ($\kB T$$>$0.2--0.5 keV) the main continuum component is well described by a power law with spectral index $\alpha_{\rm X}\sim1$ up to energies of a few hundred keV. The X-ray photons are produced by Compton up-scattering of disc UV photons by a hot electron plasma (the so-called X-ray {\it corona}). The energy loss through X-ray emission would cool down the electron plasma in a very short time scale if an efficient energy transfer mechanism from the disc to the corona were not in place. However, the physical nature of such a process is still poorly understood.

%At X-ray energies ($\kB T$$>$0.2--0.5 keV), the AGN SED presents a soft X-ray excess\footnote{An excess of soft X--ray photons that cannot be explained by simply extrapolating the power-law emission at 2--10 keV to lower energies.} (at $\kB T$$<$2 keV), a hard X--ray tail (at $\kB T$$=$2--10 keV) accompanied by a high energy cut-off at the electrons' temperature (at energies of about few hundreds of keV, \citeads{2002A&A...389..802P}), and the 6.4 keV fluorescent iron line, which are strong observational evidences for hot gas ($T\sim10^9$K) in the vicinity of the accretion disc (e.g., \citeads{1993ARA&A..31..717M}). 
%Photons coming from the accretion disc are Compton up-scattered by such hot plasma (i.e. the X--ray {\it corona}) and produce the observed hard X--ray power-law component in the AGN spectrum \citepads{1987ApJ...321..305C}. The iron emission line at 6.4 keV is usually explained by reflection of the hard X--ray photons from the underlying disc (e.g., \citeads{1989MNRAS.238..729F,1995Natur.375..659T}), while the soft X-ray excess may be originated either by a two components plasma (a soft {\it thermal} and a hard {\it non-thermal} electron distribution, \citeads{1999ASPC..161..375C}), or via blurring of emission lines due to the relativistic motion of the infalling matter (e.g. \citeads{2002MNRAS.331L..35F,2012MNRAS.420.1848D,2013MNRAS.428.2901W,2014ApJ...785...30V}).

% 
An important observational result concerning the link between the BBB (disc) and X-ray emission (corona) is provided by the non-linear correlation between the monochromatic optical luminosity at 2500 \AA\ ($\Lo$) and the one in the X--rays at 2 keV ($\Lx$, \citeads{avnitananbaum79,zamorani81,avnitananbaum86}).
The non-linear relationship between $\Lx$ and $\Lo$ (parameterised as $\lx= \gamma\lo + \beta$) has been found in both optically and X--ray selected AGN samples and exhibits a slope $\gamma$ around $0.5-0.7$ \citepads{vignali03,strateva05,steffen06,just07,green09,2010A&A...512A..34L,2010ApJ...708.1388Y,2012A&A...539A..48M,2012MNRAS.422.3268J}, implying that optically bright AGN emit relatively less X-rays than optically faint AGN. Such relation is independent of redshift, it is very tight ($\sim0.2$ dex observed dispersion, \citeads{2016ApJ...819..154L}), and it has
also been employed as a distance indicator to estimate cosmological parameters such as $\Omega_{\rm M}$ and $\Omega_\Lambda$. Thanks to the $\Lx-\Lo$ relationship (or, more precisely, its version with fluxes), \citetads{2015ApJ...815...33R} have built the first quasar Hubble diagram which extends up to $z\sim6$, in excellent agreement with the analogous Hubble diagram for Type Ia supernovae in the common redshift range (i.e. $z\sim0.01-1.4$). 
The observed $\Lx-\Lo$ correlation (or its byproduct $\aox-\Lo$, where $\aox=-0.384 \log[\Lx/\Lo]$) suggests that the disc-corona parameters are tightly linked and they must depend on the UV luminosity, and therefore ultimately on the black hole mass ($\mbh$) and accretion rate ($\dot{M}$). 
Additionally, a positive correlation between the photon index and the accretion rate has been found in several previous \revs{analyses} (e.g. \citeads{2004ApJ...607L.107W,shemmer06,2009ApJ...700L...6R}), which supports the idea that the coronal parameters \revs{are} dependent on the accretion properties onto the SMBH.
The major difficulty in interpreting such relations on physical grounds is that the origin of the X--ray emission in quasars is still a matter of debate. 

% cascade pair production
Some attempts at interpreting the X--ray spectrum of quasars (in particular its slope, the photon index $\gammax$) were based on reprocessing of radiation from a non-thermal\revs{electron}-positron pair cascade \citepads{1982ApJ...258..321S,1984MNRAS.209..175S,1983MNRAS.205..593G,1990ApJ...363L...1Z}. A highly energetic photon is absorbed in collision with an X--ray photon within the hot plasma producing an electron-positron pair, which in turn emits X-rays. This process is proportional to the source luminosity, inversely proportional to the source size, and, if the source is compact (a few tens of gravitational radii), it becomes an efficient cooling mechanism at $\kB T<100$ keV that compensates for the additional heating \citepads{1997ApJ...487..747D}. Recent results on the X-ray corona properties measured by NuSTAR have shown that pair production is an important process able to act as a thermostat in disc coronae \citepads{2015MNRAS.451.4375F}.

% adaf 
%The radiatively inefficient, optically thin, advection-dominated accretion flows (ADAF) have been proposed mainly to interpret the X--ray spectra of low-luminosity AGN (e.g. \citeads{1977ApJ...214..840I,1995ApJ...438L..37A}). However, the ADAF explanation cannot be extended to luminous quasars, since they require higher accretion rates as in SS73. 

% two phase corona model
Another possibility is provided by a two-phase accretion disc model where the entire gravitational power is dissipated via buoyancy and reconnection of magnetic fields in a uniform hot plasma immediately above (and below) the cold opaque disc \citepads{1991ApJ...380L..51H,1993ApJ...413..507H,1994ApJ...436..599S,1998MNRAS.299L..15D}. Soft X--ray photons from the disc are the main source of cooling of the hot electron within the plasma (phase 1), while the hard X-ray photons produced by the interaction via Compton scattering with the disc photons keep the X--ray corona at high temperatures (phase 2). When these two processes are balanced, a stable corona is formed. This model finds that, at a given disc temperature, the X--ray spectrum (approximated with a power-law) has a spectral index $\gammax\simeq2$, originated from a fairly compact corona, in close agreement with observations. However, this model also predicts nearly equal optical-UV and X--ray luminosities, which is not consistent with observations.
If the corona is not uniform but rather a clumpy medium, and only a fraction ($f$) of the accretion power is released in the hot phase, the resulting $\Lo/\Lx$ ratio is higher than the value computed using a model with a more uniform corona (\citeads{1994ApJ...432L..95H}). Nonetheless, the number of active blobs necessary to match observations, and the value of $f$ are\revs{relatively} arbitrary parameters.

% viscous and mhd coronae
The interaction between disc and corona has also been analysed in the viscosity-heated corona frame, where the heating in the corona is produced through friction as in the disc \citepads{2000A&A...361..175M,2002ApJ...575..117L}. However, this kind of model \revs{predicts} an overly weak corona, hence an additional source of heating is required. The importance of magnetic field turbulence has been realised not only as a supplementary heating process in the formation of the corona itself (e.g. \citeads{1979ApJ...229..318G,2002MNRAS.332..165M,2002ApJ...572L.173L}) but also as an efficient mean for the transport of the disc angular momentum (e.g. \citeads{2003ARA&A..41..555B}).
However, how the hot corona depends on $\mbh$ and $\dot{M}$ still remains an open issue.

% our approach
%Here we propose a modified version of the $\Lx-\Lo$ relation which involves the emission line full-width half maximum ($\w$), $\Lx\propto\Lo^{\hat\gamma}\w^{\hat\beta}$.
%We interpret this relationship in the context of the standard disc theory through a simple, {\it ad-hoc} model of accretion disc corona, mainly derived from the works of \citeads{1994ApJ...436..599S} and \citeads{2002MNRAS.332..165M}.
% AIM ---
%Our main aim is to provide an equation based on the observables $\Lo$, $\Lx$, and $\w$ that can be derived from a physical model linking accretion disc and X--ray corona in quasars. Our proposed relation, although very simplistic, not only shows that quasars can be considered standard candles, but it can be easily implemented in future, more sophisticated, physical models.
% -------------------------
%Powerful radio active AGN represent about 10\% of the total AGN population and show an enhanced X-ray emission linked to the jets with respect to the radio quiet ones having similar optical luminosities (e.g., \citeads{zamorani81,wilkeselvis87rl,cappi97}). Our modelling does not consider the extra complication linked to the jets, so we restrict our analysis to the radio quiet AGN only.

%The paper is structured as follows. Section...

%Source luminosities are estimated by adopting a concordance flat $\Lambda$-cosmology with $H_0=70\, \rm{km \,s^{-1}\, Mpc^{-1}}$, $\Omega_\mathrm{M}=0.3$, and $\Omega_\Lambda=0.7$ \citepads{komatsu09}.

\subsection{Outline of the work} 
In this paper we further analyse the $\Lx-\Lo$ relation with the goal of understanding its physical origin. The works summarised above suggests two main conclusions. Firstly, a tight physical relation is present between the UV-emitting disk and the X-ray emitting corona. The intrinsic dispersion of this relation is rather small, approximately 0.18 dex or less over more than four orders of magnitude \citepads{2016ApJ...819..154L}. Secondly, even though the physical connection between the accretion disc and the corona is still poorly understood, the amount of gravitational energy transferred from the disc to the corona most likely depends on the black hole mass, $\mbh$, the accretion rate, $\dot{M}$, \rev{and the black hole spin which is in turn related to the accretion radiative efficiency, $\epsilon$}. It is then straightforward to look for a physical relation connecting the X-ray luminosity with the parameters $\mbh$ and $\dot{M}$ \rev{at a given value of $\epsilon$}. If such \revs{a} relation exists, than a correspondence between $\Lx$ and $\Lo$ must be present, as $\Lo$ also depends on $\mbh$ and $\dot{M}$. We further notice that another observable relation among the same quantities is known: the virial relation between $\Lo$, $\mbh$, and the width of the broad emission lines in the optical-UV, $\w$. Summarising, we have three relations linking the two physical parameters $\mbh$ and $\dot{M}$ to three observable quantities: $\Lo=f_1(\mbh,\dot{M})$, $\Lx=f_2(\mbh,\dot{M})$, $\w=f_3(\mbh,\dot{M})$. Even without specifying the functions $f_1$, $f_2$, $f_3$, it is natural to expect that, by eliminating the parameters $\mbh$ and $\dot{M}$, we can obtain a relation between $\Lx$, $\Lo$, and $\w$.

Based on these general motivations, our approach is based on two steps:
\begin{enumerate}

\item We will investigate the $\Lx(\Lo,\w)$ relation from an observational point of view, by analysing a sample of sources with reliable measurements of the X-ray and UV luminosity and the width of a broad emission line. \revs{We will restrict our analysis to a linear} logarithmic relation, $\lx=\gamma\lo+\beta\log\w+K$. 

\item We will discuss a possible simplified model, based on the Shakura-Sunyaev accretion disc, which, through an ad-hoc assumption on the way energy is transferred from the disc to the corona, is able to reproduce the observed relation, with an exact prediction of the two slopes $\gamma$ and $\beta$ and of the normalisation $K$.  
\end{enumerate}

Powerful \rev{radio-loud} AGN represent approximately 10\% of the total AGN population and show an enhanced X-ray emission linked to the jets with respect to the radio-quiet ones having similar optical luminosities (e.g. \citeads{zamorani81,wilkeselvis87rl,cappi97,2011ApJ...726...20M}). Our toy model does not consider the extra complication linked to the jets, so we restrict our analysis to the radio-quiet AGN only.

Source luminosities are estimated by adopting a concordance flat $\Lambda$-cosmology with $H_0=70\, \rm{km \,s^{-1}\, Mpc^{-1}}$, $\Omega_\mathrm{M}=0.3$, and $\Omega_\Lambda=0.7$ \citepads{komatsu09}.

\section{The data}
\label{The Data}

The \revs{broad-lined} quasar sample we considered for our analysis has been \revs{built} following a similar approach as the one described in  \citetads{2016ApJ...819..154L} (hereafter LR16). We started with the catalogue of quasar properties presented by \citetads{2011ApJS..194...45S} \revs{(hereafter S11)}, which contains 105,783 spectroscopically confirmed broad-lined quasars. \rev{We removed from this catalogue all quasars flagged as broad absorption lines (BALs, i.e. sources with BAL\_FLAG=0 are non-BALs) and radio-loud (i.e. having a radio loudness $R=F_{\nu,6\rm{cm}}/F_{\nu,2500\AA}\geq10$, 8257 quasars, 8\% of the main SDSS sample). This yields 91,732 SDSS quasars. We further excluded 136 quasars classified as BALs by \citetads{2009ApJ...692..758G}.
This pre-cleaned SDSS quasar sample (91,596 sources) is then cross-matched with the source catalogue 3XMM--DR6} \citepads{2016A&A...590A...1R}. 3XMM--DR6 is the third generation catalogue of serendipitous X-ray sources available online and contains 678,680 X--ray source detections (468,440 unique X-ray sources) made public on or before 2016 January 31\footnote{http://xmmssc.irap.omp.eu/Catalogue/3XMM-DR6/3XMM\_DR6.html}. The net sky area covered (taking into account overlaps between observations) is $\sim$982 deg$^2$, for a net exposure time $\geq$1 ksec.
For the matching we have adopted a maximum separation of 3 arcsec to provide optical classification and spectroscopic redshift for all objects. This yields 4,303 XMM observations\revs{: 2,725 unique sources, 739 of which \revt{have} multiple ($\geq2$) observations}. 
To define a reasonably `clean' sample we have applied the following quality cuts from the 3XMM--DR6 catalogue: SUM\_FLAG$<$3 (low level of spurious detections), and HIGH\_BACKGROUND$=$0 (low background levels)\footnote{For more details the reader should refer to the 3XMM catalogue user guide at the following website http://xmmssc.irap.omp.eu/Catalogue/3XMM-DR6/3XMM-DR6\_Catalogue\_User\_Guide.html.}.
\rev{To further remove powerful radio loud} quasars we made use of the catalogue published by \citetads{2016MNRAS.462.2631M}, which is the largest available Mid-Infrared (WISE), X-ray (3XMM) and Radio (FIRST+NVSS) collection of AGN and star-forming galaxies: 2753 sources, 918 of which are considered radio-loud based on multiwavelength diagnostics (we refer to the paper for details).
We excluded 17 quasars in our sample defined as radio loud in the MIXR catalogue within a matching radius of 2 arcsec. Since the MIXR catalogue has a prior on the W3 WISE magnitude, \revs{we may be missing} a few more radio loud quasars due to \revt{that} requirement. We thus cross-matched the FIRST/NVSS catalogue (2,129,340 objects) released by \citetads{2016MNRAS.462.2631M} with the 2,708 quasars using 30 arcsec matching radius. 
\revle{We estimated the 1.4~GHz luminosity from the combined FIRST/NVSS flux values. Quasars are then classified radio-loud if they have a 1.4~GHz luminosity higher than $5\times10^{41}$ erg s$^{-1}$ (see Figures 15 and 16 by \citeads{2016MNRAS.462.2631M}).} \revt{The large majority of quasars with $L_{1.4~{\rm GHz}}>5\times10^{41}$ erg s$^{-1}$ have an $R$ ratio higher than 10.}
We found 91 objects, 23 of which are radio-loud 
%\LEt{A and A insists footnotes be incorporated into the body text when discursive and where possible.
%Please incorporate this footnote into the text }
which have been further excluded from our sample. 
We have also excluded those objects without an estimate of the 2500\AA\ flux, and the ones with a $\w$ value for the \ion{Mg}{ii} emission line lower than 2000 km s$^{-1}$ (we defer details for the latter requirement \revs{to} \S~\ref{Selection of a clean quasar sub-sample}). 

The remaining sample is composed \revt{of} 3,386 XMM observations (2,119 unique quasars with both 0.5--2 and 2--12 keV fluxes, 592 of which with two or more observations).  
Following the results presented by LR16 (see their Section~4), we decided to average all observations for sources with multiple detections that meet our selection cuts. In this way, we reduce the effect of X--ray variability on the dispersion ($\sim0.12$ dex, see \S~4 in LR16).

%------------ LUV
The $\Lo$ values are computed from the observed continuum flux densities at rest-frame 2500 \AA\ ($\Fo$) compiled by S11, which take into account the emission line contribution\footnote{These observed flux densities are divided by $(1+z)$ to shift these values into the rest-frame.}
%\LEt{A and A insists footnotes be incorporated into the body text when discursive and where possible.
%Please incorporate this footnote into the text }. %Uncertainties on $\Fo$ were not provided, we thus assumed a 2\% uncertainty on the continuum flux measurement\footnote{The average uncertainty on the bolometric luminosity values in the SDSS quasar catalogue is $\sim$3\%. Since the bolometric measurements have an additional uncertainty due to the bolometric correction employed, we considered 2\% a reasonable value for the uncertainties on the continuum monochromatic fluxes.}. 
%------------

%------------ LX
The X--ray monochromatic luminosities and the $5\sigma$ minimum detectable flux at 2 keV ($\fmin$) are estimated following a similar approach to the one described by LR16 (we refer to their Section~2.1 for details). 
Specifically, the X-ray monochromatic luminosities are estimated from the soft 0.5-2~keV and 2-12~keV fluxes, $F_S$ and $F_H$, respectively, in the 3XMM-DR6 catalogue. These fluxes are computed assuming in each band a power law spectrum with photon index $\gammax$=1.7. We then used a single power law to fit $F_S$ and F$_H$ simultaneously, obtaining a new value of the photon index, $\Gamma_{\rm X,1}$. Since the distribution of $\Gamma_{\rm X,1}$ is relatively broad (extending from $\sim1$ to $\sim3$ with an average of 1.9) and the effective area of the EPIC instruments has a strong energy dependence, a different photon index implies slightly different flux estimates in the two bands. We therefore apply a correction based on the average effective area as a function of energy, and estimate two corrected fluxes $F_{S,1}$, $F_{S,2}$. Then we repeat the power law fit and estimate a new photon index, $\Gamma_{\rm X,2}$. We iterate this procedure until we obtain a convergence ($\Delta \gammax <0.01$). The monochromatic flux at 2~keV (rest frame) is obtained from the final power law.
%A small fraction of sources have uncertainties on the X--ray fluxes at 2 keV smaller than 2\%, which may be underestimated. We have thus considered a minimum 3\% uncertainty on $\Lx$. 
%------------
For each detected object we have also computed the EPIC sensitivity ($5\sigma$ minimum detectable flux) at 2 keV using the same procedure as in LR16. Briefly, we have considered the pn, MOS1, and MOS2 on-time and off-axis values, where both MOS1 and MOS2 on-time and off-axis have been combined. The total MOS on-time and off-axis are \revs{the largest and smallest values} of the two individual cameras, respectively. 
We then estimated the minimum detectable flux in the soft band corrected for vignetting. 
\revle{The sensitivity flux values at 2 keV ($\fmin$) are then calculated by assuming a photon index of \revs{1.9} and finally combined by taking the sum of the pn and MOS fluxes in the case where both values are available.}
%\LEt{A and A insists footnotes be incorporated into the body text when discursive and where possible.
%Please incorporate this footnote into the text }. 

The catalogue of SDSS-DR7 quasar properties published by S11 also contains a wealth of additional information that we considered in our analysis, such as, for example, black hole mass estimates, Eddington ratios, and $\w$ for several lines, along with their uncertainties. We thus refer to S11 for details on how these parameters have been computed.

\subsection{Selection of a{ `clean'} quasar sub-sample}
\label{Selection of a clean quasar sub-sample}
Our aim is to select a sub-sample covering a redshift range as wide as possible, with precise estimates of $\Lo$ and $\Lx$, by removing systematic effects and low-quality measurements. The possible sources of contamination/{systematic error} are: dust reddening in the optical-UV, gas absorption in X-rays, large statistical errors due to low X-ray flux, and {\it Eddington bias} due to the flux limit of the X-ray observations. Here we briefly discuss each of these points, and describe the filters we applied to obtain the final `best' sample for our analysis.
\begin{enumerate}
\item Dust reddening. We used the SDSS photometry to compute, for each object, the slope $\Gamma_{1}$ of a $\log(\nu)-\log(\nu L\nu)$ power law in the rest frame 0.3--1 $\mu$m range, and the analogous slope $\Gamma_{2}$ in the 1450--3000~\AA\ range.  
\rev{The wavelength coverage of the SDSS photometry does not cover the rest frame $1\mu m$ at $z\sim2$, we thus considered the mid-infrared data from {\it Spitzer} and WISE, and the near-infrared data from the Two Micron All Sky Survey and UKIDSS published by \citetads{2013ApJS..206....4K} for the same SDSS-DR7 quasar sample.}
We selected all sources in the ($\Gamma_{1}$, $\Gamma_{2}$) plane within a circle centred at the reference values for a standard SED of quasars (see \citeads{2015ApJ...815...33R} and LR16 for further details) with a radius corresponding to a reddening $E(B-V)\simeq0.1$. 
\rev{Ninety-three percent of the sample (1965/2119) have $E(B-V)\leq0.1$.}

\item Signal-to-Noise. We selected only sources with a S/N$>$5 in the full 0.2--12~keV EPIC band. We did not apply an analogous filter on the optical spectra because we expect the error on the continuum flux measurement is always small compared with the uncertainties in the X-rays, for all the quasars in the spectroscopic SDSS samples. \rev{The fraction of objects excluded by such a signal-to-noise cut is 10\%.}

\item X-ray absorption. In order to avoid systematic underestimates of the the X-ray flux, we selected only sources with an X-ray photon index $\gammax>1.6$. Furthermore, we also excluded a few objects with $\gammax>3.0$. This filter is needed to avoid strong outliers (95\% of our objects have $1.6<\gammax<2.8$) which may be due to observational issues such as incorrect background subtraction in one band, and/or low S/N.

\item Eddington bias. Due to X-ray variability, sources with an average X-ray flux close to the flux limit of the observation will be observed only in case of a positive flux fluctuation. This introduces an ``Eddington bias'' towards high fluxes. 
We therefore minimised this bias by including only sources whose minimum detectable flux is lower than the expected X-ray flux for each observation (we refer to Appendix~A in LR16). \rev{On average, we have that the minimum detectable monochromatic flux at 2 keV is approximately $3\times10^{-32}$ erg s$^{-1}$ cm$^{-2}$ Hz$^{-1}$. However, we caution that this value should not be considered the `survey limiting flux' since the 3XMM--DR6 catalogue is not a proper flux-limited sample, but rather a collection of all XMM observations performed in a given period. It is thus not trivial to estimate the expected minimum flux for the whole data set.}
The effects of this bias can be reduced if\revs{no}-detections are included in the statistical analysis. However, this would complicate the statistical analysis, and would make the estimate of the intrinsic dispersion of the correlations much more uncertain. \rev{LR16 showed that there is no significant variation on both slope and intercept (within their uncertainties) among censored and detected samples once the Eddington bias is minimised using the method discussed in their Appendix~A. We therefore decided not to include censored data in our current work.}
We note that our procedure to minimise the Eddington bias is slightly circular: we need the $\Lx-\Lo$ relation in order to estimate the `expected' X-ray flux. We assumed that the expected slopes 0.6 with a variable normalisation. This filter is applied iteratively in a similar way as discussed in LR16. We obtained a perfect convergence (i.e. \revt{no} more sources rejected) after just one iteration.

\item Homogeneous line properties. In order to explore the dependence among $\Lx$, $\Lo$, and $\w$, we selected only sources with an observation of the \ion{Mg}{ii} $\lambda$2800~\AA\ full-width half maximum higher than 2000 km s$^{-1}$.  This filter ensures homogeneity in the estimate of the line parameters, and, given the redshift distribution of our sample, \revs{rejects only 20\% of the sample which lies outside the range where \ion{Mg}{ii} is observed ($z<0.35$ and $z>2.3$), and 5\%} of the sample with $\w<2000$~km s$^{-1}$ within that redshift range. \revle{This provides the \revs{widest} redshift coverage possible given the SDSS spectral coverage. Other emission lines as, for example, \ion{H}{$\beta$} and \ion{C}{iv} would bias the sample towards low/high redshift quasars, thus are less suitable for cosmological studies. It is also well known that the \ion{C}{iv} line may be affected by non-gravitational motions (e.g. outflows), and therefore the interpretation for that line is likely to be non trivial.}
%\LEt{A and A insists footnotes be incorporated into the body text when discursive and where possible.
%Please incorporate this footnote into the text }.
\end{enumerate}
While all the points discussed above have a clear qualitative justification, it is not obvious how to choose the numerical parameters of each filter based on general considerations. Here we apply a very conservative empirical criterion: if the effects discussed above introduce a bias in the estimates of the X-ray and UV fluxes, we will observe a change in the $\Lx-\Lo-\w$ relation when we start applying a given filter. There will be a range in the filter parameters where if we tighten the filter, the bias is reduced, and the measured relation changes. When the bias is completely removed, if we further tighten the filter, we will not see any further change in the relation, but only a decrease in the number of selected sources. 
The parameters described above are all significantly beyond these threshold values: in other words, if we relax some of them (for example, allowing for more reddened objects, or accepting sources with $\gammax>1.5$, or changing the threshold for the Eddington bias from 2~$\sigma$ to 1.5~$\sigma$) this has no effect on the results of the analysis of the $\Lx-\Lo-\w$ relation. 

\begin{figure}
 \centering
  \includegraphics[width=\linewidth,clip]{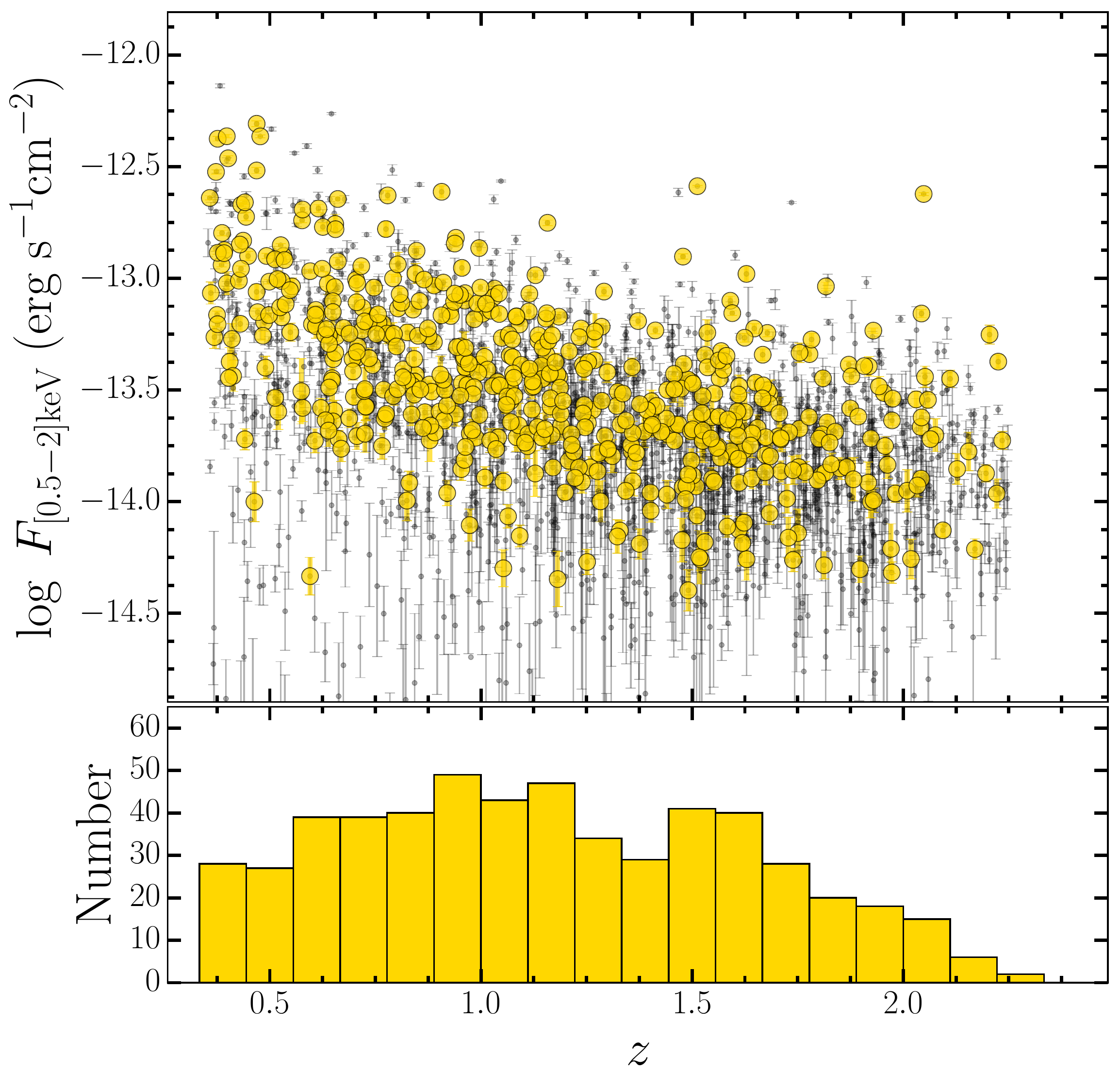}
  \caption{Top panel: Distribution of the observed $0.5-2$keV flux versus redshift for the main sample of 2,119 quasars (small black points) and the one of 545 sources (gold points). Bottom panel: Redshift distribution for the best quasar sample.}
  \label{fluxSz}
\end{figure}

The final ``best'' sample consists of 545 sources and is shown in Figure \ref{fluxSz}, spanning a redshift range of 0.358--2.234. This is only $\sim$25\% of the initial sample. As mentioned above, we could relax the filtering criteria and add a few hundred more objects, but at the price of less precise measurements and possible residual small systematic effects. Since our aim is to analyse a three-parameter linear correlation, we believe our sample is large enough for our purpose, and we prefer to favour sample cleanness over higher statistics. This choice is supported by the results of the analysis, which are discussed in detail in Section~\ref{Statistical analysis}: the errors on the parameters of the $\Lx-\Lo-\w$ correlation are relatively small;  relaxing the parameters and adding more sources (still in the bias-free range) means we obtain a larger intrinsic dispersion (due to the lower quality measurements included in the sample) and the final uncertainties on the best fit parameters do not decrease significantly. 

%-------------------------------------------------------------
%-------------------------------------------------------------
%-------------------------------------------------------------
\begin{figure}
 \centering
  \includegraphics[width=\linewidth,clip]{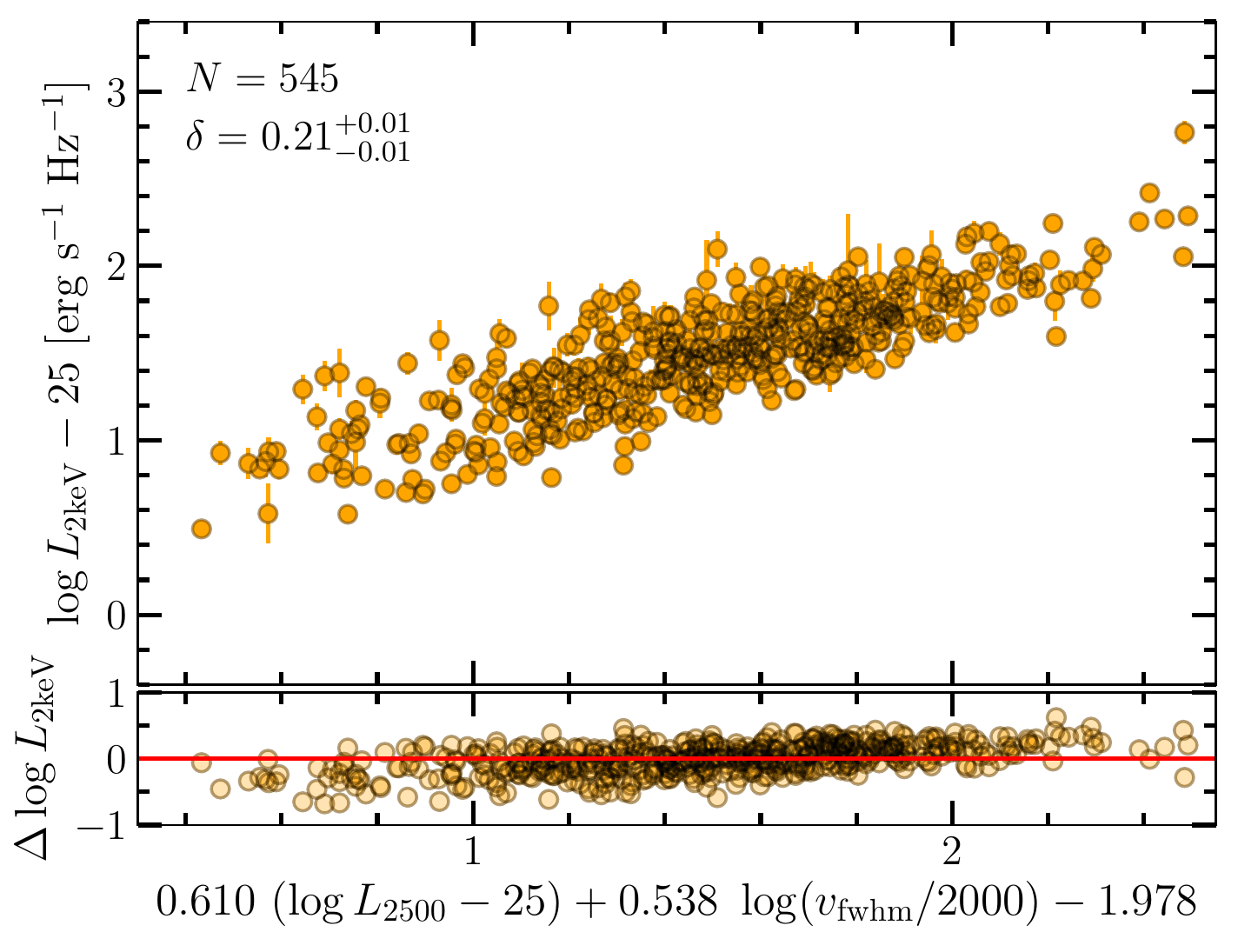}
  \caption{The $\Lx-\Lo-\w$ plane seen edge-on. Both optical and X-ray luminosities have been normalised to $10^{25}$ erg s$^{-1}$ Hz$^{-1}$. The $\w$ values are taken from S11 and are normalised to 2000 km s$^{-1}$. Here we considered the $\w$ estimated from the whole \ion{Mg}{ii} emission line. The bottom panel shows the residuals defined as the difference between the observed monochromatic X-ray luminosity at 2 keV and the best fitting function.}
  \label{lxlowall}
\end{figure}
%-------------------------------------------------------------
\section{Statistical analysis}
\label{Statistical analysis}

To fit the data in the three-dimensional plane defined as $\log\Lx=\hat{\gamma}\log\Lo+\hat\beta\log\w+\hat{K}$, we adopted the Python package {\it emcee} \citepads{2013PASP..125..306F}, which is a pure-Python implementation of Goodman \& Weare's affine invariant Markov chain Monte Carlo (MCMC) ensemble sampler. We accounted for the dependent variable $Z$ against the independent variables $X$ and $Y$ (i.e. $Z=\hat{\gamma} X + \hat{\beta} Y + \hat{K}$), and we included uncertainties on both $\Lx$ and $\w$. 
\revle{Since uncertainties on $\Lo$ are not provided in the SDSS quasar catalogue, we assumed a 2\% uncertainty on the continuum flux measurement, which is roughly the average uncertainty on the bolometric luminosity values in the SDSS quasar catalogue.} % is roughly $\sim$3\%. Since $L_{\rm bol}$ values have an additional uncertainty due to the bolometric correction considered, we employed 2\% as a reasonable value for the uncertainties on the continuum monochromatic fluxes at 2500\AA.%\LEt{A and A insists footnotes be incorporated into the body text when discursive and where possible. Please incorporate this footnote into the text }. 
\revt{None of the results presented significantly depend on our assumed uncertainty on the optical data.}
\revt{Slope and intercept estimated assuming 2\% optical uncertainty are consistent within 1.5$\sigma$ with the fits performed with 5\% and 10\% uncertainty on $\Lo$, and a normal distribution of uncertainties on $\Lo$ having a mean at $\sim$0.25 dex ($\sim30$\%).} 
We then fitted $\hat{\gamma}$, $\hat{\beta}$, $\hat{K}$, and the intrinsic dispersion of the relation $\delta$.
We normalised $\Lo$ and $\Lx$ to $10^{25}$ erg s$^{-1}$ Hz$^{-1}$, while $\w$ has been normalised to 2000 km s$^{-1}$. Here we considered the $\w$ values obtained from the whole \ion{Mg}{ii} emission line. \citetads{2011ApJS..194...42R} found that their \ion{Mg}{ii} line dispersion estimates are rather different from those of \citetads{2008ApJ...680..169S}. Specifically, they found that using $\w$ as a line width indicator may overestimate the $\mbh$ values for the broadest emission lines, and underestimate $\mbh$ for the narrowest emission lines as compared with $\mbh$ computed using the line dispersion. This may introduce a tilt of the plane in the $\w$ direction. 
Even though we believe that the $\w$ estimates provided by S11 are more robust than those published in their previous analysis. Without other measurements for a statistically significant sample of quasars to compare with, we cannot explore this point any further.

%-------------------------------------------------------------
\begin{figure}
 \centering
  \includegraphics[width=\linewidth,clip]{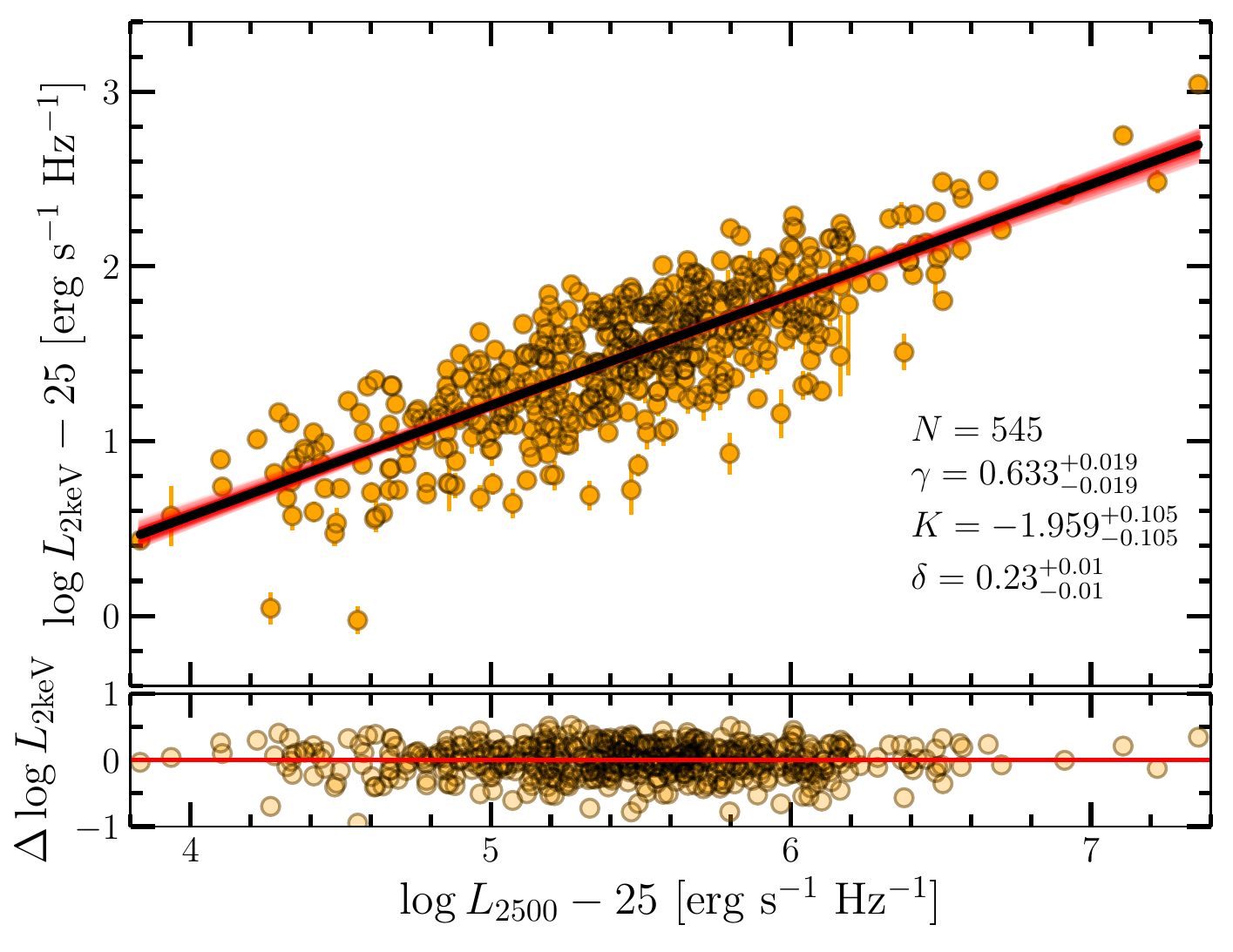}
  \caption{Rest-frame monochromatic luminosities $\lx$ against $\lo$ for the final ``best'' sample of 545 quasars (orange circles) as described in \S~\ref{Selection of a clean quasar sub-sample}. The results from the emcee regression (black solid line) are also reported. Red thin lines represent 100 different realisations of the $\Lo-\Lx$ relation. The lower panel shows the residuals of $\lx$ and $\lo$ with respect to the emcee best-fit line.}
  \label{lxloall}
\end{figure}
%-------------------------------------------------------------

%%%%%%%%%%%%%%%%%%%%%
\begin{table*}                                                          
\caption{Results of the regression analysis of the $\Lx-\Lo-\w$ plane parametrised as $\log \Lx = \hat{\gamma} \log \Lo + \hat{\beta} \log \w + \hat{K}$. \label{tbl-1}}
\centering                     
\begin{tabular}{lccccc}      
\hline\noalign{\smallskip}                
Sample & $\hat\gamma$  & $\hat\beta$   & $\hat{K}$ & $\delta$ & $N_{\rm tot}$\tablefootmark{a} \\
\hline\noalign{\smallskip}
 Main                    & 0.564$\pm$0.014   &   0.379$\pm$0.046   &   -1.725$\pm$0.076 & 0.32$\pm$0.01 & 2119(592)  \\    
 E(B--V)$\leq$0.1 & 0.562$\pm$0.015   &   0.404$\pm$0.048   &   -1.722$\pm$0.083 & 0.31$\pm$0.01 & 1965(551)  \\   
 E(B--V)$\leq$0.1 -- S/N$>$5 & 0.556$\pm$0.015 & 0.389$\pm$0.048 & -1.662$\pm$0.082 & 0.30$\pm$0.01 & 1758(459) \\
 E(B--V)$\leq$0.1 -- S/N$>$5 -- $1.6\leq\gammax\leq2.8$ & 0.559$\pm$0.014 & 0.499$\pm$0.048 & -1.660$\pm$0.076 & 0.24$\pm$0.01 & 1345(292) \\
 E(B--V)$\leq$0.1 -- S/N$>$5 -- $1.9\leq\gammax\leq2.8$ & 0.579$\pm$0.015 & 0.552$\pm$0.053 & -1.789$\pm$0.084 & 0.22$\pm$0.01 & 929(179) \\  
 E(B--V)$\leq$0.1 -- S/N$>$5 -- $1.9\leq\gammax\leq2.8$ -- EB\tablefootmark{b} & 0.610$\pm$0.019 & 0.538$\pm$0.072 & -1.978$\pm$0.100 & 0.21$\pm$0.01 & 545(105) \\  
\hline\noalign{\smallskip} 
\end{tabular}                                  
\tablefoot{
\tablefoottext{a}{The total number of fitted quasars. The value between parentheses refers to the number of quasars with multiple observations within the sample.}
\tablefoottext{b}{EB= Eddington Bias correction applied.}
}                                 
\end{table*}                                                                       
%%%%%%%%%%%%%%%%%%%

To check if our results depend on the regression technique, we also considered the LINMIX\_ERR\footnote{This algorithm has been implemented in Python and its description can be found at http://linmix.readthedocs.org/en/latest/src/linmix.html.} method \citepads{2007ApJ...665.1489K}, which takes account \revs{of} measurement uncertainties on both independent and dependent \revs{variables}, non-detections (not considered in the following analysis), and intrinsic scatter by adopting a Bayesian approach to compute the posterior probability distribution of parameters, given the observed data.
In general, the resulting fitting parameters output of LINMIX\_ERR are fully consistent within the uncertainties of those of emcee, thus we decided to report the parameter values from emcee only.

\revs{ 
Figure~\ref{lxlowall} shows an edge-on view of the $\Lx-\Lo-\w$ plane. The best-fit regression equation is
\begin{align}
\label{lxlowemcee}
(\log\Lx - 25) = (0.610\pm0.019)(\log \Lo -25) + \nonumber\\(0.538\pm0.072)[\log \w - (3+\log2)] +(-1.978\pm0.100),
\end{align}
with a dispersion of $\sim$0.21 dex, which is an extremely tight correlation (with a Pearson $R$ coefficient of 0.83)%\LEt{Please check that I have retained your intended meaning.}. 
The Student's t-test for $\hat\gamma$ and $\hat\beta$ yields a significance that these slopes are different from zero of approximately $32$ and $7\sigma$, respectively. 
}
We have also evaluated how slopes and dispersion change as a function of various selection cuts. 
The summary of our findings is provided in Table~\ref{tbl-1}. Interestingly, both slopes are roughly consistent within the 2$\sigma$ level irrespective of the selection cut, yet the dispersion decreases as the cuts become more conservative on the quality of the data.
%Additionally, we have analysed the $\Lx-\Lo-\w$ relation on the sample quasars with multiple observations only (105 objects) finding $\hat\gamma=0.603\pm0.036$, $\hat\beta=0.329\pm0.139$, and $\hat{K}=-1.866\pm0.197$ with a fitted dispersion $\delta=0.18\pm0.01$ dex. 
A more in-depth analysis of the dispersion and of how it depends on X--ray variability, disc inclination, and so on is the subject of a forthcoming paper.
Figure~\ref{lxloall} shows the $\Lx-\Lo$ relation without the dependence on $\w$ \revs{($R=0.81$), with $\gamma$ statistically different from zero at $\sim32\sigma$.} Both slope and normalisation are in agreement with previous works in the literature on optically selected samples (e.g. \citeads{strateva05,steffen06}; LR16). These findings are not altered if we repeat the analysis by doubling the uncertainties on $\Lx$.  
\revs{The Fisher's F-test\footnote{\revs{The $F$ parameter is defined as $F=\frac{{\rm d}f(R^2-R_r^2)}{m(1-R^2)}$ where $m=2$, $R_r$ is the Pearson coefficient of the $\Lx-\Lo$ relation, $R^2$ and d$f$ are the values of the Pearson coefficient and residuals for the $\Lx-\Lo-\w$ relation.}} %\LEt{A and A insists footnotes be incorporated into the body text when discursive and where possible. Please incorporate this footnote into the text }
for the additional dependence of the $\Lx-\Lo$ relation on $\w$ yields $F=28$ with a $p-$value of $2.7\times10^{-12}$ that the dependence on $\w$ is not significant.}
We must notice that the residuals ($\Delta\log\Lx$, defined as the difference between the observed monochromatic X-ray luminosity at 2 keV and the best fitting function) plotted in Figure~\ref{lxlowall} are slightly tilted, meaning that the best fit under/overestimate\revt{s} the slope at low/high values of $\Lo$ and $\w$.
To investigate this, further we evaluated how slopes and dispersion in the $\Lx-\Lo-\w$ plane change by slicing the $\Lx-\Lo$ and the $\Lx-\w$ relationships as a function of $\w$ and $\Lo$, respectively. \revs{Details are provided in Appendix~\ref{appendixa}. Summarising, we found that the additional dependence of the observed $\Lx-\Lo$ correlation on $\w$ is statistically significant and may point towards a much tighter connection between accretion disc physics and X-ray corona than has been found previously.}

\subsection{Redshift dependence}
\label{Redshift dependence}
%-------------------------------------------------------------
\begin{figure}
 \includegraphics[width=\linewidth,clip]{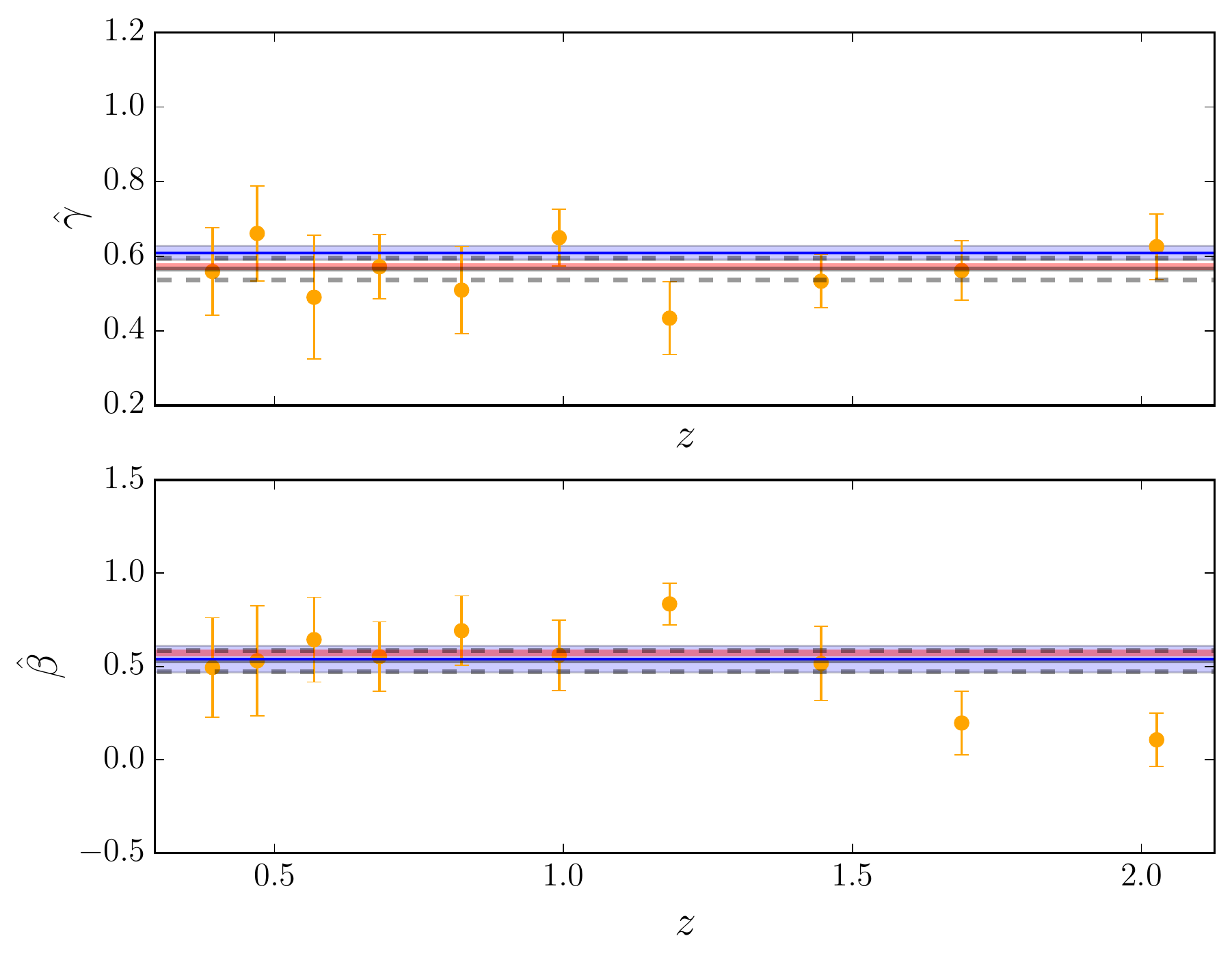}
 \caption{Evolution of the $\hat\gamma$ and $\hat\beta$ parameters with redshift. Orange points (along with their uncertainties) are the results of the fitting procedure of the function $\fx = \hat\gamma_z\fo+\hat\beta_z\w+\hat{K}$ in narrow redshift bins (see \S~\ref{Redshift dependence} for details).
 The red solid line represents the predicted value of $4/7$ (\S~\ref{The model}). The blue solid line and shaded area represent the $\hat\gamma$ and $\hat\beta$ from the global fit using luminosities as discussed in \S~\ref{Statistical analysis}. The grey solid and dashed lines are the weighted means with uncertainties of the orange points.}
 \label{gammabetaz}
\end{figure}
%-------------------------------------------------------------
We now \revs{discuss} the redshift evolution $\Lx-\Lo-\w$ relation where fluxes are considered instead of luminosities. To do so, we \revs{analysed} the best quasar sample only as this will be the one used to compute the cosmological parameters in Paper II.
\rev{We divided our sample into equally spaced redshift bins in $\log z$ with a $\Delta\log z <0.1$ to minimise the scatter due to the different luminosity distances compared to the intrinsic dispersion within each bin. We also require sufficiently populated bins (i.e. $N\geq20$ objects) for a statistically meaningful fit.
We split the sample into ten intervals with $\Delta\log z =0.08$ and, for each redshift interval, we performed a fit of the $\Fx-\Fo-\w$ relation, $\fx = \hat\gamma_z\fo+\hat\beta_z\w+\hat{K}$, with free $\hat\gamma_z$ and $\hat\beta_z$. The results for the best-fit slopes $\hat\gamma_z$ and $\hat\beta_z$ as a function redshift are shown in Figure~\ref{gammabetaz}. 
Both $\hat\gamma_z$ and $\hat\beta_z$ slopes are consistent with a constant value at all redshifts, without any significant evolution with time, and best-fit weighted mean values of $\hat\gamma_z\simeq 0.566\pm0.030$ and $\hat\beta_z\simeq0.529\pm0.056$.}
%The relation between fluxes is thus not solely a good proxy of that between luminosities, but also both slopes are statistically consistent with the predicted ones ($\gamma=\beta=4/7$) within $1\sigma$. 

\section{Toy model}
\label{The model}
In this section we outline a simple, ad-hoc physical model which provides the relationships between the disc and the X-ray emission in terms of the black hole mass, the accretion rate, and the distance to the black hole ($R$). Our main goal is to link such relations with observable quantities, thus delivering a correlation that can be easily compared with the results of our statistical analysis.
In our modelisation, we introduce the following non-dimensional parameters:
\begin{align}
\label{nodimpar}
m = \frac{\mbh}{M_\odot}, \quad r=\frac{R}{\Rs}=\frac{R}{3\times10^5 {\rm cm}} m^{-1}, \nonumber\\
 \dot{m}=\frac{\dot{M}}{\dot{M}_{\rm Edd}} \simeq \frac{\dot{M}}{10^{18} {\rm gr~s^{-1}}}  m^{-1} ,
\end{align}
where $\Rs=2G\mbh/c^2$ is the Schwarzschild radius, $\dot{M}_{\rm Edd}$ is the Eddington accretion
\begin{equation}
\label{eddacc}
\dot{M}_{\rm Edd} = \frac{\epsilon L_{\rm Edd}}{c^2},
\end{equation}
and $L_{\rm Edd} = \frac{4 \pi c G \mbh \mu m_{\rm p} }{\sigmaT}$ defines the Eddington luminosity: $\sigmaT$ is the Thomson cross section, $\mu$ is the mean atomic weight, which we assumed to be equal to 0.615 throughout the paper (approximate for a fully ionised cosmic mixture), and $\epsilon$ is the radiative efficiency assumed to be 1/12 (appropriate for a steady SMBH).
%To compute the monochromatic luminosity at a given frequency we assume that $L_\nu \propto \nu^{-\gamma_{\rm o}}$, and thus
%\begin{equation}
%\label{monlum}
%L_\nu = \frac{L_{\rm c} ~(1-\gamma_{\rm o})}{\nu } \left( \frac{\nu}{\nu_{\rm c}} \right)^{1-\gamma_{\rm o}},
%\end{equation}
%where $L_{\rm c}$ and $\nu_{\rm c}$ are the characteristic (total) luminosity and frequency, respectively.
%Equation (\ref{monlum}) is a reasonable description of the disc radiation spectrum at the frequency of interest in the optical, yet it cannot be used to describe the monochromatic emission in the X--rays since $\gammax\simeq2$ and its energy band width is an essential factor. To compute the  X--ray monochromatic luminosity, we considered a power-law in the frequency interval $\nu_1-\nu_2$. We thus have
%\begin{equation}
%\label{monlum_range}
%L_\nu = L_{\rm \nu_2-\nu_1,c} (1-\alpha_{\rm X}) \frac{\nu^{-\alpha_{\rm X}}}{\nu_2^{1-\alpha_{\rm X}} - \nu_1^{1-\alpha_{\rm X}} } ,
%\end{equation}
%where $L_{\rm \nu_2-\nu_1,c}$ is now the X--ray characteristic emission in the $\nu_2-\nu_1$ range,
%and $\alpha_{\rm X}$ represents the slope of the X--ray SED (related to the X--ray photon index as $\alpha_{\rm X} = \gammax - 1$). The formula for the monochromatic X--ray luminosity thus depends on the frequency range chosen. 
%If $\alpha_{\rm X}=1$, equation (\ref{monlum_range}) reduces to $L_\nu = L_{\rm \nu_2-\nu_1,c}~\nu^{-1}/\ln(\nu_2/\nu_1)$.

To compare the results with observations, we need to have a relation that transforms our physical variables $m$ and $\dot{m}$ to the observable ones. 
The parameter $m$ is usually obtained from the {\it virial} relationship (\citeads{2014SSRv..183..253P}, and references therein)
\begin{equation}
\label{mvir}
m = \frac{R_{\rm blr} \w^2}{G M_\odot},
\end{equation}
\revle{where $\w$ is the full width at half maximum of the line, $R_{\rm blr}$ represents the size of the broad line region, and a virial coefficient factor, which accounts for our ignorance of the broad line region geometry, of the order of unity \citepads{2004MNRAS.352.1390M}.}
%\LEt{A and A insists footnotes be incorporated into the body text when discursive and where possible. Please incorporate this footnote into the text }.
%This factor can assume different values depending on the calibrations. For instance, f=1 in McLure & Dunlop (2004) and f\sim 5.5 (e.g., Onken et al. 2004) in Vestergaard & Peterson (2006).a factor $f$ is a dimensionless factor of the order of unity}.
The parameter $R_{\rm blr}$ is proportional to the square root of the quasar bolometric luminosity ($L_{\rm bol}$), $R_{\rm blr}= k L_{\rm bol}^{0.5}$ \citepads{2015JKAS...48..203T}, and $\kappa$ is a calibration constant. From equation~(\ref{mvir}) we have that 
\begin{equation}
\label{mdef}
m  \simeq 7.5\times 10^{-27} \kappa L_{\rm bol}^{0.5} \w^2 \simeq 3 \times 10^{13} \kappa L_{\rm bol,46}^{0.5} \wtwo^2,
\end{equation}
where $L_{\rm bol,46}$ and $\wtwo$ are normalised to $10^{46}$ erg s$^{-1}$ and 2000 km s$^{-1}$, respectively.

\subsection{The predicted monochromatic luminosity at 2500 \AA}
%Accretion onto the SMBH is a very complicated problem. 
The most widely used model of accretion disc physics is provided by the disc structure equations published by \citetads{1973A&A....24..337S} (hereafter SS73), which assume hydrostatic equilibrium in the vertical direction, conservation of angular momentum (the torque vanishes at the inner stable circular orbit), and disc energy balance. The exchange of angular momentum along the disc is provided by viscous stresses within the fluid, through an ad-hoc effective viscosity, and making use of standard hydrodynamics without the inclusion of magnetic fields. This approach has been very effective especially in simulations, since the full magnetohydrodynamic treatment can be numerically costly. For these reasons, the SS73 parametrisation still finds wide application today (see \citeads{2013LRR....16....1A} for a detailed discussion. \revle{On the other hand, from an observational perspective, the SS73 model has been challenged, especially in reproducing the shape of the SED near the ultraviolet peak \citepads{2012MNRAS.423..451L} through the soft X-ray region (e.g. \citeads{2001ApJ...553..987B}). Moreover, it does not provide an explanation for the X--ray emission observed in quasars at energies higher than 0.5 keV.}
%\LEt{A and A insists footnotes be incorporated into the body text when discursive and where possible. Please incorporate this footnote into the text }).
%%%%%%%%%%%%%
In what follows, we considered \revs{an approach similar to that in \citetads{2003MNRAS.341.1051M}}. We \revs{summarise} below the basic assumptions and equations that we use throughout the paper.

The total disc luminosity ($L_{\rm disc}$) of a Keplerian geometrically thin, optically thick, accretion disc is obtained from the integral of the viscous dissipation rate (e.g. \citeads{1981ARA&A..19..137P,1993ApJ...403...94F}), at a given $R$, within the disc defined as
\begin{equation}
\label{dr}
D(R) =  \frac{3GM \dot{M}}{4 \pi R^3} \left[ 1- \left(\frac{R_0}{R}\right)^{1/2}\right],
\end{equation}
where $\left[ 1- \left(\frac{R_0}{R}\right)^{1/2}\right]$ is a dimensionless factor set by the inner boundary condition at the radius $R_0$. \revs{When written as above, this factor does not} include relativistic effects \citepads{1973blho.conf..343N}, which are important but not dominant (we refer to equations 3--5 and table 1 in \citeads{2011MNRAS.417..681L}, and \citeads{1995ApJ...450..508R,2003MNRAS.342..951M} for a detailed discussion). In what follows we thus assume a Newtonian regime.
From equation (\ref{dr}) the expression of $L_{\rm disc}$ %from $R_0$ to a certain distance $R$ is
%\begin{align}
%\label{ldisc_def}
%L_{\rm disc} (R_0,R)= 2\pi \int^R_{R_0} R D(R) {\rm d}R = \nonumber \\  \frac{GM\dot{M}}{2R_0} - \frac{3GM\dot{M}}{2R}\left[ 1- \frac{2}{3}\left(\frac{R_0}{R}\right)^{1/2}\right].
%\end{align}
%We can neglect the second term in the equation (\ref{ldisc_def}) at large radii. By considering the inner radius of a non-rotating SMBH 
at $R_0=3\Rs$ (i.e. the innermost stable circular orbit of a non-rotating SMBH) is %and $R>>R_0$, we finally have that 
\begin{equation}
\label{ldisc}
L_{\rm disc} =  \frac{G M \dot{M}}{6\Rs} = \frac{\dot{M} c^2}{12} \simeq 7.7\times10^{37} ~ m~ \dot{m} \quad {\rm erg~s^{-1}}.
\end{equation}
If each element of the disc radiates as a black body, the disc temperature is given by equating the black body flux to the dissipation rate. The disc temperature at a given distance can be written as 
\begin{equation}
\label{eqt}
T_{\rm disc}(r) = T_0 ~r^{-3/4} J(r)^{1/4}
%\simeq 2.7\times10^7 m^{-1/4} ~\dot{m}^{1/4}~{\rm K},
,\end{equation}
where $J(r)=( 1 - \sqrt{3/r})$ and $T_0=(3GM\dot{M}/8\pi\sigmaB\Rs^3)^{1/4}$.
We assume for simplicity that there are no additional dissipative mechanisms within the disc, which will produce soft photons that will interact with \revs{those} in the plasma, lowering the temperature through the Compton cooling mechanism.

\rev{By assuming that $L_\nu \propto \nu^{-\gamma_{\rm o}}$ we can now estimate the monochromatic optical luminosity at 2500~\AA\ as}
\begin{equation}
\label{monlumoptdef}
\Lo = \frac{L_{\rm disc}~(1-\gamma_{\rm o})}{\nuo} \left( \frac{\nuo}{\nu_{\rm c}} \right)^{1-\gamma_{\rm o}},
\end{equation}
where $\nu_{\rm c}$ is the characteristic frequency of the disc, and $\nuo$ is the reference optical frequency. The peak frequency of the big blue bump is
\begin{equation}
\label{nuc}
%\nu_{\rm c} \simeq 2.8 \frac{\kB}{h} T_{\rm disc} \simeq 4.7\times10^{18} m^{-1/4} ~\dot{m}^{1/4}~ r^{-3/4}\quad \rm Hz.
\nu_{\rm c} \simeq 2.8 \frac{\kB}{h} T_{\rm disc}(r_{\rm max}) \simeq 7.2\times10^{17} m^{-1/4} ~\dot{m}^{1/4}\quad \rm Hz,
\end{equation}
where $T_{\rm disc}(r_{\rm max})$ is defined at the radius where the maximum in the disc temperature occurs (i.e. $r_{\rm max}=49/12$, thus $T_{\rm disc}(r_{\rm max})\simeq0.21 T_0\simeq1.2\times10^7 m^{-1/4} ~\dot{m}^{1/4}$).
The final expression for $\Lo$ is 
\begin{align}
\label{monlumopt}
%\Lo \simeq \frac{1.3\times10^{38}}{\nuo^{\gamma_{\rm o}}~(1-\gamma_{\rm o})} \left(1.3 \times 10^{18}\right)^{(\gamma_{\rm o}-1)} \times \nonumber \\ m^{(5-\gamma_{\rm o})/4} ~\dot{m}^{(3+\gamma_{\rm o})/4}~ r^{-(3\gamma_{\rm o}+1)/4} \quad {\rm erg~s^{-1}~Hz^{-1}}.
\Lo \simeq \frac{7.7\times10^{37}~(1-\gamma_{\rm o})}{\nuo^{\gamma_{\rm o}}} \left(7.2\times10^{17}\right)^{(\gamma_{\rm o}-1)} \times \nonumber \\ m^{(5-\gamma_{\rm o})/4} ~\dot{m}^{(3+\gamma_{\rm o})/4},
\end{align}
in erg s$^{-1}$ Hz$^{-1}$.
For a SS73 disc, the optical-ultraviolet spectrum has $\gamma_{\rm o}=-1/3$, which has been proved to be \revs{consistent} through observations of the polarised light {interior to} the dust-emitting region in several low-redshift quasars \citepads{2008Natur.454..492K}.
We finally have that 
\begin{equation}
\label{lo}
\Lo\simeq 1.7\times10^{19} m^{4/3}~\dot{m}^{2/3}$ erg s$^{-1}$ Hz$^{-1}, 
\end{equation}
which expresses the dependence of $\Lo$ on the physical parameters $m$ and $\dot{m}$. %For $\mbh=10^8M_\odot$ and maximum accretion, the predicted $\Lo$ is $2.6\times10^{29}$  erg s$^{-1}$ Hz$^{-1}$, and it reduces to $5.6\times10^{28}$  erg s$^{-1}$ Hz$^{-1}$ for $\dot{m}=0.1$, in good agreement with observations.

\subsection{The predicted monochromatic luminosity at 2 keV}
\label{The monochromatic X--ray luminosity}
Our formalism is mainly based on the one in \citetads{1994ApJ...436..599S} (SZ94 hereinafter). They discussed the physical conditions of the disc-corona system under the assumption that the cold accretion disc is geometrically thin (i.e. SS73). In their model, a fraction ($f$) of the accretion power associated with the transport of angular momentum stored in the disc (via magnetic field loops, and/or flux tubes, for instance) is dissipated (only once the accretion power reaches the disc surface) in a hot corona through an unspecified physical dissipation mechanism. 
The accretion rate in the disc is thus $(1-f)$ of the total accretion. 
SZ94 found that for $f\simeq1,$ the emission from a hot corona becomes relevant only when the gas pressure dominates over radiation inside the disc. The radius ($r_{\rm tr}$) defining the boundary between the radiation dominated region and the gas pressure dominated zone \revs{(i.e. the outer/inner boundary of the radiation/gas pressure dominated zone)} can be computed by assuming that the radiation pressure equates the gas pressure, and thus
\begin{equation}
\label{rtr}
r_{\rm tr} \simeq 120 \left( \alpha m \right)^{2/21} \dot{m}^{16/21} (1-f)^{6/7} J(r)^{16/21},
\end{equation}
% point out that the size at rtr is small consistent with observations.
%Rapid variability of the 2-10 keV X-ray emission seen from many AGN indicates that the corona is physically small. 
in units of $\Rs$,
%\LEt{A and A insists footnotes be incorporated into the body text when discursive and where possible.Please incorporate this footnote into the text }
\revle{where $\alpha$ is the standard disc viscosity parameter. We note that equation (\ref{rtr}) assumes a radiative diffusion of the order of unity (i.e. SS73), and a normalisation factor that differs from equation (29) in SZ94, as the latter has been estimated for $\epsilon=1$ and $\mu=1$.
Additionally,} the $r J(r)^{-16/21}$ factor has a minimum at $\tilde{r}=5.72$, which corresponds to a critical accretion rate of the order of 1\%. For $\dot{m}$ lower than the critical one, the factor $120 \left( \alpha m \right)^{2/21} \dot{m}^{16/21} (1-f)^{6/7}$ is always smaller than $r J(r)^{-16/21}$ and there are no solutions (see SZ79 and \citeads{2002MNRAS.332..165M} for additional discussion). Therefore, our toy model is valid only for accretion rates higher than the critical one.
To estimate the X--ray coronal emission we assume that the hot corona is powered by a magnetised flux emerging (in a \revs{dissipationless} fashion) from the disc to balance the amplifying effects of the magneto-rotational instability (MRI), which produces turbulence able to transport angular momentum in the disc itself. A detailed discussion of this phenomenon is presented by \citetads{2002MNRAS.332..165M} (hereafter MF02, see their Section 2 and references therein). MRI works at full efficiency in cases where the disc is gas-pressure dominated. When the disc is radiation-pressure dominated, the MRI is quenched and the (much lower) value of the emerging disc flux becomes, to a certain extent, model-dependent.
We assume a sharp transition at the radius $r_{\rm tr}$ where the gas and radiation pressures are equal. In the gas dominated zone, the total power released locally by the accretion is divided equally between the power radiated by the disc itself and \revs{that} transferred to the corona. On the other side (i.e. the radiation dominated zone), all the power is radiated within the disc and the amount of energy transferred to the corona is negligible. Therefore, given our assumptions, we have that 
\begin{equation}
\label{lcor}
L_{\rm cor} \simeq 2 \times 10^{36} \alpha^{-2/21} m^{19/21} \dot{m}^{5/21} (1-f)^{-6/7} J(r)^{-16/21} \quad {\rm erg~s^{-1}}.
\end{equation}

%------------------------------------------------- 
%\begin{figure}
% \includegraphics[width=\linewidth,clip]{lx_mbh_analysis}
% \caption{ }\label{lx_mbh}
%\end{figure}
%------------------------------------------------- 
The monochromatic X--ray luminosity at 2 keV can be written as
\begin{equation}
\label{monlumxdef}
\Lx = L_{\rm cor} (1-\alpha_{\rm X}) \frac{\nux^{-\alpha_{\rm X}}}{\nu_2^{1-\alpha_{\rm X}} - \nu_1^{1-\alpha_{\rm X}} },
\end{equation}
where $\alpha_{\rm X}$ represents the slope of the X--ray SED (related to the X--ray photon index as $\alpha_{\rm X} = \gammax - 1$). 
We considered an energy range between 0.1 keV and 40 keV, which has a mean geometric energy of 2 keV (i.e. $\sqrt{0.1 \times 40}\simeq 2$ keV). 
By substituting equation (\ref{lcor}) into the above equation, and assuming a $\alpha_{\rm X}=0.9$ \citepads{2009ApJS..183...17Y,2010ApJ...708.1388Y}, we finally have
\begin{equation}
\label{lx}
\Lx \simeq 6.6 \times 10^{17}~\alpha^{-2/21} m^{19/21} \dot{m}^{5/21} ~(1-f)^{-6/7}~ J(r)^{-16/21}
~ \quad {\rm erg~s^{-1}}.
\end{equation}
%For $\mbh=10^8M_\odot$ maximum accretion, and $\alpha=1$, the predicted $\Lx$ is $2.2\times10^{25}$  erg s$^{-1}$ Hz$^{-1}$. The $\aox$ value for such $\mbh$ is $\sim -1.6$, remarkably consistent with the results based on quasar samples mainly selected from SDSS\footnote{The $\aox$ distribution observed in optically selected quasars typically cover the range (-1.2,-1.8), with a mean value of about -1.6 (\citeads{strateva05,steffen06,just07,green09}).}.

%The effect of modifying the energy range and/or $\alpha_{\rm X}$ alters the normalization on equation~(\ref{lx}).
%The maximum energy we considered on equation~(\ref{monlumxdef}) corresponds to a coronal temperature of $5\times10^8$ K. If we instead consider a higher coronal temperature, as for example $10^9$ K, the corresponding pivot point is at 3 keV, not too far from the energy of our interest. With the latter $T_{\rm cor}$ value, the normalization on equation~(\ref{lx}) will be reduced by 14\%. By employing $\alpha_{\rm X}=2.1$ and $T_{\rm cor}=5\times10^8$ K the normalization becomes $2 \times 10^{17}$, a factor of 3.5 lower than the one on equation~(\ref{lx}).

\subsection{The predicted relation between X--ray and optical luminosity of quasars}
\begin{figure}
 \includegraphics[width=\linewidth,clip]{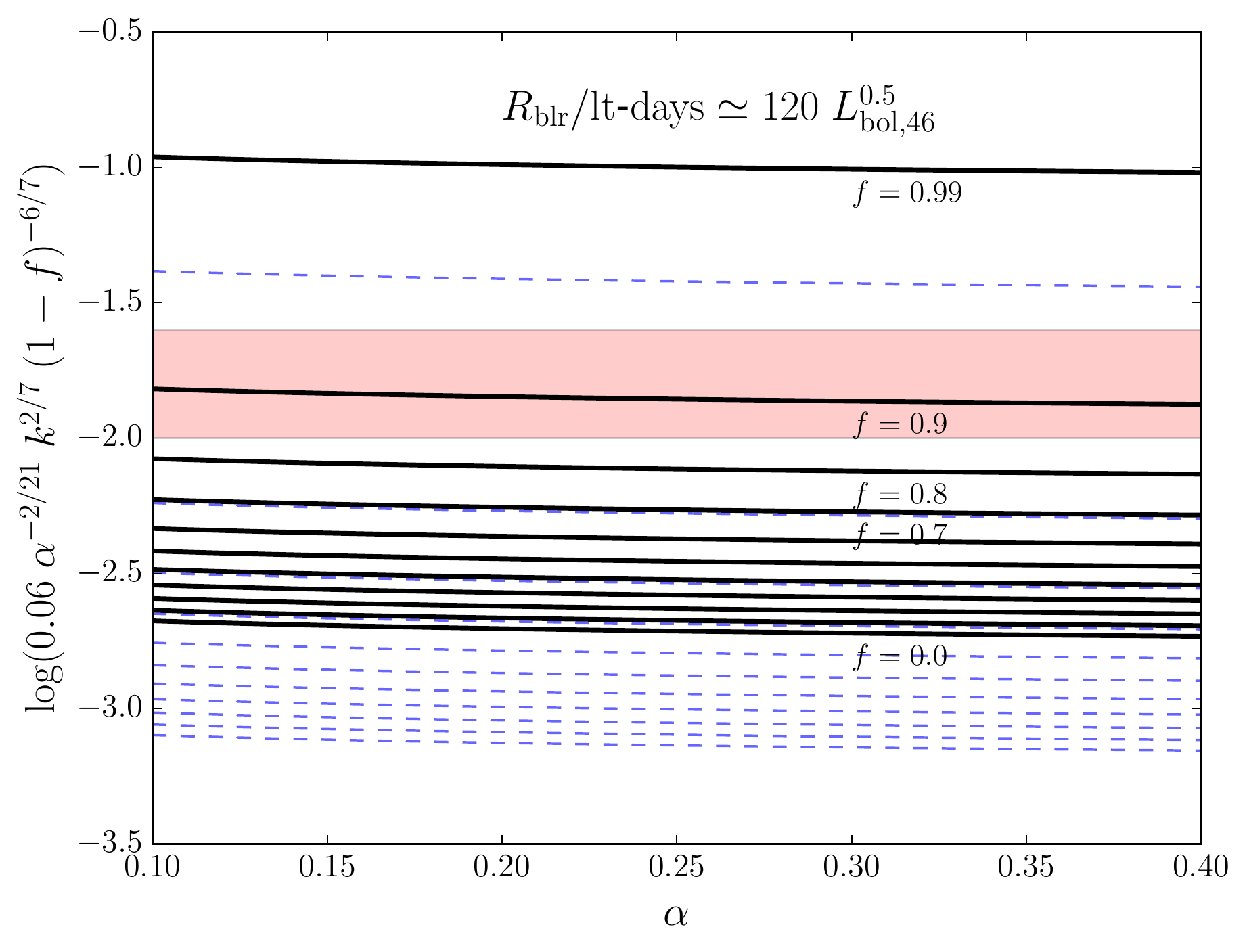}
 \caption{\rev{The logarithm of the normalisation of equation~(\ref{lxlownorm}) as a function of $\alpha$. The $\kappa$ factor has been fixed to $3\times10^{-6}$, corresponding to a broad line region size of approximately $3\times10^{17}$ cm ($\sim120$ light days) for a bolometric luminosity of $10^{46}$ erg s$^{-1}$.} The solid black lines correspond to different values of $f$, which go from 0 to 0.99 from bottom to top. The dashed blue lines correspond to the normalisation at $\sim$4 light days in the same range of $f$. The red shaded area indicates the range of observed $\hat{K}$ (see table~\ref{tbl-1}).}
 \label{normlolx}
\end{figure}

The physical relation between $\Lx$ and $\Lo$ can be calculated by solving the system of equations (\ref{mdef}), (\ref{lo}), and (\ref{lx}). 
Provided that $L_{\rm bol}$ is given by equation~(\ref{ldisc}), we can write equation~(\ref{mdef}) as
\begin{equation}
m \simeq 4.3\times10^{-15} \kappa^2~ \dot{m} ~ \w^4.
\end{equation}
%\begin{equation}
%\label{w}
%\w \propto m^{1/4}~ \dot{m}^{-1/4}.
%\end{equation}
It follows that
\begin{align}
\label{lxmw}
%\Lx \propto m^{\frac{24}{21}} \w^{-\frac{20}{21}} J(r)^{\frac{16}{21}},
\Lx \simeq 6.6\times10^{4}\dot{m}^{24/21}\alpha^{-2/21}(1-f)^{-6/7} J(r)^{-16/21} \kappa^{38/21} \w^{76/21}
\end{align}
and
\begin{equation}
\label{lomw}
%\Lo \propto m^{2} \w^{-(3+\gamma_{\rm o})} r^{-(3\gamma_{\rm o}+1)/4}.
\Lo \simeq 1.2~\dot{m}^2 \kappa^{8/3} \w^{16/3}
\end{equation}
in erg s$^{-1}$.
From the two equations above we finally have 
%\begin{equation}
%\label{lxlowgeneral}
%\Lx \propto \Lo^{\frac{4}{7}} \w^{\frac{(16+12\gamma_{\rm o})}{21}} r^{\frac{3 (\gamma_{\rm o}-1)}{7}} J(r)^{-\frac{16}{21}},
%\end{equation}
%which provides the dependence of $\Lx$ on the observables $\Lo$ and $\w$. Assuming an SS73 optical-UV disc emission slope, $\gamma_{\rm o}=1/3$, equation (\ref{lxlowgeneral}) becomes 
\begin{align}
\label{lxlow}
\Lx \simeq 6\times10^4~\Lo^{4/7}~\w^{4/7}~\alpha^{-2/21}  \kappa^{2/7} \times \nonumber \\(1-f)^{-6/7}J(r)^{-16/21} ~ {\rm erg~s^{-1}}.
\end{align}
To compare the relation above with the results of our regression analysis, we must normalise both $\Lx$ and $\Lo$ to $10^{25}$ erg s$^{-1}$ Hz$^{-1}$ and $\w$ to 2000 km s$^{-1}$, thus equation (\ref{lxlow}) becomes
%We have now a physical relation based on the observables $\Lo$, $\Lx$, and $\w$ that we can compare with the data.
\begin{equation}
\label{lxlownorm}
L_{{\rm X},25} \simeq ~0.06 ~L_{{\rm UV},25}^{4/7}~\upsilon_{{\rm fwhm},2000}^{4/7}~\alpha^{-2/21}  \kappa^{2/7} (1-f)^{-6/7} J(r)^{-16/21}.
\end{equation}
We note that the additional correction due to the factor $J(r)^{-16/21}$ of equation~(\ref{lxlow}) makes the slopes from $\hat{\gamma}=\hat{\beta}=0.571$ to $\hat{\gamma}\simeq0.581$ and $\hat{\beta}\simeq0.531$\footnote{Here we assumed that $r~J(r)^{-16/21}$ can be roughly approximated with $r^\delta$ in equation~(\ref{rtr}).}, still in agreement with our findings.
Figure~\ref{normlolx} shows how the normalisation of equation~(\ref{lxlownorm}) changes as a function of $\alpha$, $\kappa$, and $f$. \rev{We fixed the $\kappa$ factor to $3\times10^{-6}$, which corresponds to a broad line region size of approximately $3\times10^{17}$ cm ($\sim120$ light days, or $\sim$0.1 parsec) for a bolometric luminosity of $10^{46}$ erg s$^{-1}$, typical of quasars (e.g. \citeads{2005ApJ...629...61K})}. The $\alpha$ parameter varies in the range 0.1--0.4 \citepads{2007MNRAS.376.1740K}, while $f$ goes from 0 to 0.99. The normalisation of the relation presents a large scatter, which can be a factor of $\sim2$ in logarithm (i.e. orders of magnitude) depending on the value of $f$ (see also Fig.~1 in \citeads{2003MNRAS.341.1051M}). 
Despite our crude assumptions, not only are the observed slopes of the $\Lx-\Lo-\w$ relation in really good agreement with the ones of equation~(\ref{lxlownorm}), but also the observed normalisation is in a range consistent with the predictions.

%-------------------------------------------------------------
%-------------------------------------------------------------

\begin{figure}
 \includegraphics[width=\linewidth,clip]{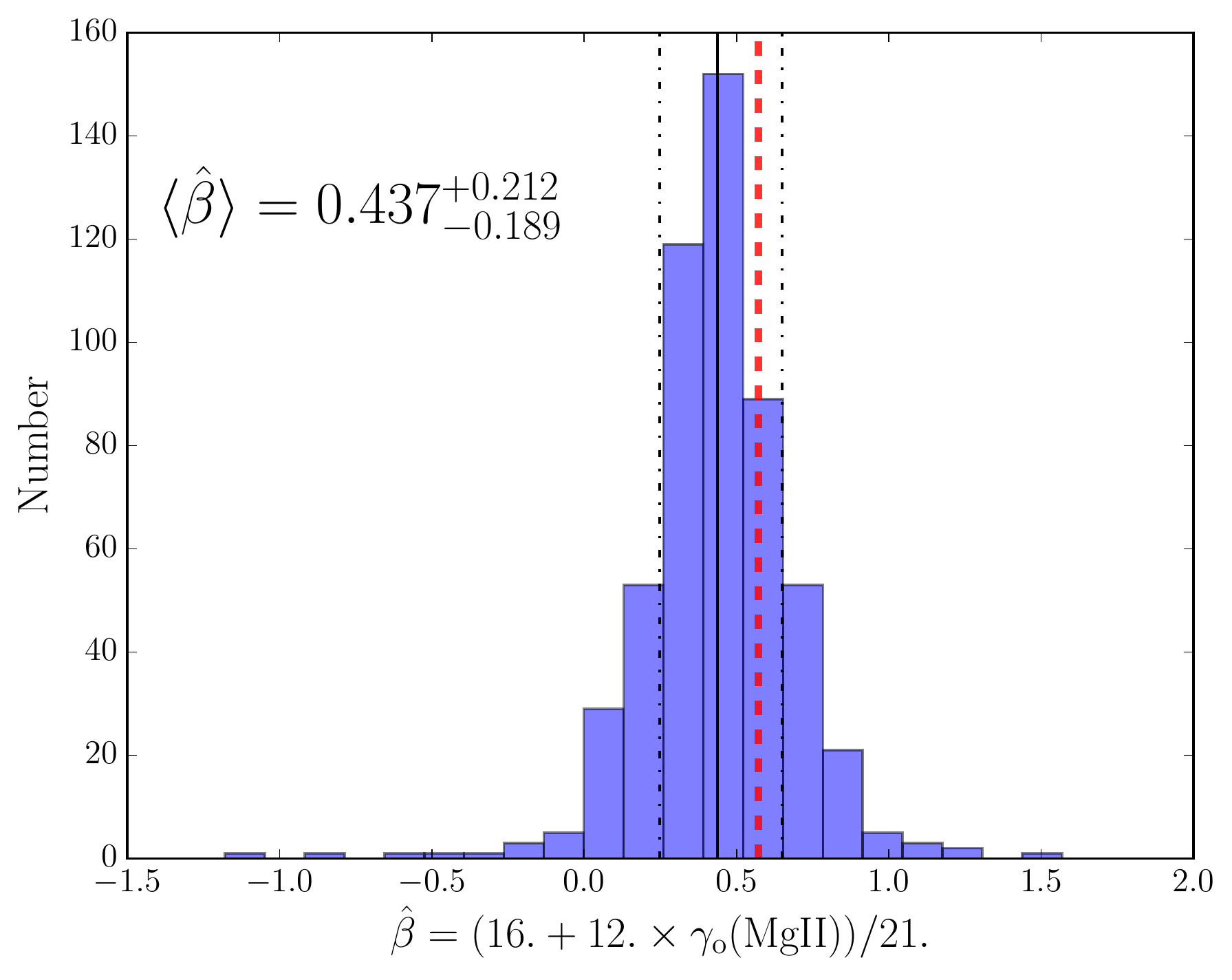}
 \caption{\rev{The $\hat\beta$ parameter as a function of the continuum optical slopes estimated for the \ion{Mg}{ii} emission line. The red dashed line represents $\hat\beta$ for $\gamma_{\rm o}=-1/3$. The solid and dot-dashed lines represent the median, the 16th and the 84th percentile, respectively.}}
 \label{w_betaslope}
\end{figure}

\rev{We have also investigated how the slopes vary if we take $\gamma_{\rm o}$ as a free parameter. Equation~(\ref{lxlow}) becomes
\begin{equation}
\Lx\propto\Lo^{4/7}\w^{(16+12\gamma_{\rm o})/12},
\end{equation}
thus $\hat{\gamma}$ does not have any dependence on $\gamma_{\rm o}$. By considering the continuum spectral slopes estimated by S11 for the \ion{Mg}{ii} emission line, we can explore how $\hat{\beta}$ varies as a function of $\gamma_{\rm o}$. Such slopes are estimated by taking the continuum+iron fitting windows in the range [2200,2700] \AA\ and [2900,3090] \AA. The pseudo-continuum is then subtracted from the spectrum, and the \ion{Mg}{ii} emission line is fitted over the [2700,2900] \AA\ wavelength range using different sets of Gaussians (we defer to their section 3.3 for further details). Figure~\ref{w_betaslope} presents the $\hat\beta$ parameter as a function of the estimated continuum optical slopes discussed above. The red dashed line represents $\hat\beta$ for $\gamma_{\rm o}=-1/3$. The average $\hat\beta$ is 0.437$^{+0.212}_{-0.189}$, consistent within 1$\sigma$ level with the one estimated for the clean quasar sample by fitting the $\Lx-\Lo-\w$ plane.

}

%-------------------------------------------------------------
%-------------------------------------------------------------

\section{Discussion}
\label{Discussion}

\subsection{The choice of observables}
\label{Observables}
% justify the choices of the monochromatic wavelength 2500 A and X-ray energy 2 keV. 
Previous monochromatic luminosity correlation studies between the X-ray and optical  have employed the energy at 2 keV and the wavelength at 2500 \AA, respectively. Such preference was principally due to historical reasons: the first analyses back in the early 1980s were carried out with data where these monochromatic values were easy to measure \citepads{avnitananbaum79,zamorani81,avnitananbaum86}.
Largely used even nowadays (principally for the sake of comparison with previous works), such endpoints at 2500 \AA\ and 2 keV are still observationally convenient. \revs{In fact, the 2500 \AA\ is not an overly red wavelength to be contaminated by the host galaxy and it is not too blue to be absorbed by neutral hydrogen.}
The 2 keV flux is also a convenient choice, considering that the 0.5--2 keV integral has the highest effective area in the main X-ray telescopes, and that 2 keV is high enough to be insensitive to Galactic absorption.

The possible dependence of the $\Lx-\Lo$ relation on the choice of the X--ray and UV frequencies has been explored by 
\citetads{2010ApJ...708.1388Y}, based on a sample of 327 quasars with SDSS optical spectra and XMM--{\it Newton} X-ray data. No significant change in the relation has been found for different choices of $\nu_{\rm UV}$ and $\nu_{\rm X}$.
%is a more arbitrary choice, thus other options should be explored\footnote{The soft-excess observed in a significant number of radio quiet AGN contaminates energies roughly between 0.1 keV and 1 keV, hence energies higher than 2 keV may be a more sensible alternative.}. 

%By cross-correlating the SDSS DR5 quasar catalogue with the XMM-{\it Newton} archive, \citetads{2010ApJ...708.1388Y} have studied the $\aox-\Lo$ relation at different optical and X--ray energies (i.e. 1500, 2500, and 5000 \AA\, and 1, 1.5, 2, 4, 7, or 10 keV) for 327 quasars where both optical and X-ray spectra were available. They found that the choice of optical wavelength does not strongly influence the $\aox-\Lo$ relation, while it becomes steeper when $\aox$ is defined at low ($\sim$1 keV) X--ray energies, and marginally flatter at high ($\sim$10 keV) X-ray energies with respect to the standard 2 keV option.

\citetads{2012MNRAS.422.3268J}  presented a novel method to investigate this point further by defining the ``optical to X-ray correlation spectrum'' based on a new ``correlation spectrum technique''. Briefly, the X--ray luminosity in a specific energy band is cross-correlated (via the Spearman's rank test) with the monochromatic luminosity at each wavelength of the optical spectrum in each single object of their sample (51 optically selected quasars from SDSS DR7).
This interesting technique delivers a more systematic and statistically robust comparison between the X--ray and optical portion of the quasar spectrum (we refer to their figure~2), but requires very high X--ray spectral quality: minimum 2000 counts in at least one XMM-{\it Newton} EPIC band. They found that the correlation coefficient between the 2--10 keV luminosity and the optical spectra is rather flat over the whole spectral range covered by SDSS, meaning that the 2--10 keV luminosity equally strongly correlates with every single wavelength point in the SDSS spectrum.

For the purposes of our analysis we thus decided to consider the standard 2500 \AA\ and 2 keV monochromatic luminosities, although we believe that a more systematic analysis with a larger quasar sample is required. 

The additional observable required in our study is a proxy for the virial velocity, which can be either the line dispersion or the full width at half maximum of the line. \citetads{2011ApJS..194...42R} pointed out that the line dispersion may be a better proxy than $\w$ for the virial velocity, although the line dispersion is more sensitive to the wings of the line profile \citepads{2006A&A...456...75C}. 
Given that currently there is no consensus on which version of the calibration ($\w$ versus line dispersion) for the virial velocity is better, we decided to employ the $\w$ measurements which are estimated for statistically significant large quasar samples.

\subsection{Implications on the coronal size}
The main shortcoming of our toy model is that the bulk of the coronal emission is produced at $r_{\rm tr}$, which can be rather large. For example, the transition radius for $m=10^{8-9}$ with a typical accretion rate of $\dot{m}=0.1-0.3$ and $\alpha=0.4$ ranges from a few to several hundred $\rg$, depending on the value of $f$ (the higher $f$, the smaller $r_{\rm tr}$). 
\rev{
Such an assumption is relatively crude and is not meant to provide any prediction on the possible geometry of the corona surrounding the disc, but it is rather an ad-hoc postulation to obtain a reasonable relation between observables compared with the data. Even if in our toy model the overall radiation budget of the corona is estimated at $r_{\rm tr}$, the effective location of the corona may also be placed at smaller radii.
From a qualitative perspective, magnetic reconnection can also arise from the torque (caused by rotation) of magnetic field lines (originated at $r\geq r_{\rm tr}$) that bridge two opposite sides of the accretion disc. This may cause particle acceleration at the magnetic reconnection site, located closer to the SMBH (at $r < r_{\rm tr}$), where particles could lose their energy radiatively via interactions with the surrounding radiation field (e.g. via inverse Compton).}

As a zero order comparison, models where magnetic reconnection events give rise to the X--ray emission (the so-called ``flare'' models) consider coronal scales on the order of a few tens up to hundreds of gravitational radii (see for example \citeads{2004A&A...420....1C}, \citeads{2011A&A...530A.136T}). 
\rev{In a growing number of cases, the X-ray reverberation lags and the emissivity profile of the quasar accretion disc (i.e. the illumination pattern of the disc by the X-rays emitted from the corona) suggest a corona that extends at low height over the surface of the disc itself (e.g. \citeads{2013MNRAS.430..247W,2015MNRAS.448..703W} and \citeads{2016MNRAS.458..200W}).
Additionally, in the recent work by \citetads{2016AN....337..441D}, a very compact X--ray source model has been questioned as it is extremely difficult for a nearly point-like object above the black hole to intercept sufficient seed photons from the disk in order to generate the hard X--ray Compton continuum, which in turn produces the observed iron line/reflected spectrum in quasars.
}
%From an observational perspective, the coronal size measured in the lensed quasar Q 2237$+$0305 is found to be roughly 40 $\rg$ \citepads{2013ApJ...769...53M}.  
%We should also point out that, in our simple modelling, the presence of the corona has the effect to change the disc parameters, i.e. it increases of the disc density and optical depth, and it reduces the disc height. Consequently, the \textit{actual} transition radius may be significantly smaller. 
\rev{
We conclude that the actual extent of the X-ray source still remains a matter of debate, as there are\revs{few} sources where this measurement is available, and that any further comparison of our toy model with measures of X--ray coronal sizes would require additional physics, which is beyond the scope of this paper}.

\subsection{Implication for black hole mass and broad line region calibrators}
One of the basic assumptions we made when calculating equation~(\ref{lxlow}) is that the the size of the broad line region scales with the bolometric emission, rather than the continuum luminosity underlying a specific emission line. Currently, single epoch black hole mass estimators are based on the ad-hoc hypothesis that a reasonable proxy of $R_{\rm blr}$ is the monochromatic luminosity ($\lambda L_\lambda$) of a given emission line raised to a certain power ($R_{\rm blr}\propto (\lambda L_\lambda)^{0.5}$). This assumption is also present in reverberation mapping studies of the $R_{\rm blr}-L$ relation \citepads{2007ApJ...659..997K,2009ApJ...697..160B}.
% maybe find more ref here...
However, a systematic discrepancy exists between black hole mass values (and thus $R_{\rm blr}$) computed with different emission lines calibrators (e.g. \citeads{2005ApJ...629...61K}).
Such discrepancy may be due to both the virial estimator considered and the exact prescription used for the line characterisation (e.g. \citeads{2009ApJ...692..246D}).

We thus explored to what degree equation~(\ref{lxlow}) changes if we consider the ionising photons at a certain wavelength instead of $\lbol$ (i.e. $m\propto \w^2~\Lo^{0.5}$). By doing this we find that $\hat{\gamma}$ does not change, while $\hat\beta$ becomes 6/7, which is an overly steep slope compared with our observations. Our toy model thus  suggests that the radius of the quasar broad line region is likely to scale with the bolometric luminosity rather than $\lambda L_\lambda$ \citepads{2015JKAS...48..203T}.

\section{Conclusions}
The tight $\Lx-\Lo$ relationship in quasars has a dispersion of $\sim$0.24~dex over approximately five orders of magnitude in luminosity (e.g. LR16), which indicates that there is good ``coupling" between the disk, emitting the primary radiation and the hot-electron {\it corona,} emitting X--rays. 
This is the observational evidence that a specific physical mechanism must regulate the energy transfer from the disc to the corona. Additionally, the non-linearity of such a relation provides a novel, powerful way to estimate the quasar absolute luminosity, making these objects {\it standard candles}.
Here we present a modified version of the $\Lx-\Lo$ relation which involves the emission line full-width half maximum ($\w$), $\Lx\propto\Lo^{\hat\gamma}\w^{\hat\beta}$. 
\rev{The homogeneous sample used here has a dispersion of $\delta\sim0.21$ and, based on the analysis presented in \citetads{2016ApJ...819..154L}, we expect that this dispersion can be further reduced if precise, well calibrated X-ray observations are performed. In LR16 we estimated an intrinsic dispersion not higher than $\delta\simeq0.18-0.19$, consistent with the one estimated with the best quasar sample with multiple observations only. In fact, our cuts have the effect of selecting preferentially detections with low off-axis angle ($<$20--25) and sources with low {equivalent width} of the [\ion{O}{iii}] $\lambda$\revs{5007}\AA\ line ($<50$, i.e. nearly face-on, we refer to \citeads{2011MNRAS.411.2223R,2017MNRAS.464..385B}).}

We also interpreted such a relation through a toy (but physically motivated) model based on the ones presented by \citetads{1994ApJ...436..599S} and \citetads{2002MNRAS.332..165M}, where a geometrically thin, optically thick accretion disc with a magnetised uniform corona are discussed. % in the context of the observed relation between $\Lo$ and $\Lx$ in quasars.
We assumed that the corona is mainly powered by the accretion disc at the transition radius ($r_{\rm tr}$) where the gas equates the radiation pressure. %The corona thus has a {\it doughnut}-like structure, with lower emission in the inner disc region where the radiation is dominant, while the bulk of the coronal radiation will be in the gas dominated zone at (or near) $r_{\rm tr}$.
We find a coronal luminosity depending on the physical parameters $\mbh$ and accretion rate $\dot{M}$ as $L_{\rm cor} \propto \mbh^{19/21} (\dot{M}/\dot{M}_{\rm Edd})^{5/21}$. 
We then estimated the monochromatic optical-UV and X--ray luminosities at 2500 \AA\ and 2 keV as $\Lo\propto \mbh^{4/3} (\dot{M}/\dot{M}_{\rm Edd})^{2/3}$ (for $\gamma_{\rm o}=1/3$) and $\Lx\propto L_{\rm cor}$ (with a normalisation factor dependent on the value of $\gammax$).
Assuming a broad line region size function of the bolometric luminosity $R_{\rm blr}\propto L_{\rm bol}^{0.5}$ we have that $M_{\rm bh} \propto \dot{M}/\dot{M}_{\rm Edd} \w^4$, which leads to the final relation $\Lx\propto\Lo^{4/7} \w^{4/7}$. Such relation is remarkably consistent with the fit obtained from a sample of 545 optically selected quasars from SDSS DR7 cross-matched with the latest XMM--{\it Newton} catalogue 3XMM-DR6. %The toy model we presented, although very crude, is capable stringent \revs{of making highly constrained predictions} on the X--ray luminosities (at a given optical-UV emission and $\w$) of blue quasars.
The toy model we presented, although very crude, \revs{provides interesting predictions of how the X-ray luminosity varies as a function of both the optical emission and $\w$ for unobscured/blue quasars.}
%Such good agreement with the observations also implicitly confirms that $R_{\rm blr}$ scales with the bolometric radiation, rather than the monochromatic emission of a given line.
We have also shown that the proposed relation $\Lx\propto\Lo^{4/7} \w^{4/7}$ does not evolve with time \revs{(in the redshift range covered by our data, $0.358<z<2.234$)}, and thus it can be employed as a cosmological indicator to robustly estimate cosmological parameters (e.g. $\Omega_{\rm M}$, $\Omega_{\Lambda}$, $w_0$, and $w_a$). 
%Our analysis finally prove that the optical-UV/X-ray emission of blue luminous quasars is powered by a common physical mechanism, which makes such sources {\it standard candles}.

\begin{acknowledgements}
We deeply thank the referee for his/her useful comments and suggestions which have significantly improved the clarity of the paper.
We are grateful to Marco Salvati for giving us the basis of the toy model presented here and for insightful discussion. 
E.L. thanks Andrea Merloni for careful reading the paper and providing useful comments and Beatrix Mingo for clarifications on the use of her published MIXR catalogue. 
E.L. also thanks the University of Keele for the kind hospitality while writing this paper.
E.L. is supported by a European Union COFUND/Durham Junior Research Fellowship (under EU grant agreement no. 609412).
This work has been supported by the grants PRIN-INAF 2012 and ASI INAF NuSTAR  I/037/12/0.
For all catalogue correlations we have used the Virtual Observatory software TOPCAT \citepads{2005ASPC..347...29T} available online (http://www.star.bris.ac.uk/$\sim$mbt/topcat/).
This research has made use of data obtained from the 3XMM XMM--Newton serendipitous source catalogue compiled by the ten institutes of the XMM--Newton Survey Science Centre selected by ESA. This research made use of matplotlib, a Python library for publication quality graphics \citepads{2007CSE.....9...90H}.
\end{acknowledgements}
%%%%%%%%%%%%%%%%%%%%%%%%%%%%%%%%%%%%%%%%%%%%%%%%%%



\bibliographystyle{aa}
\bibliography{bibl} % if your bibtex file is called example.bib

%%%%%%%%%%%%%%%%%%%%%%%%%%%%%%%%%%%%%%%%%%%%%%%%%%

%%%%%%%%%%%%%%%%% APPENDICES %%%%%%%%%%%%%%%%%%%%%

\appendix
\section{Additional statistical analysis of the $\Lx-\Lo-\w$ relation.}
\label{appendixa}
%-------------------------------------------------------------
\begin{figure*}[t!]
 \subfloat{\includegraphics[width=0.45\linewidth,clip]{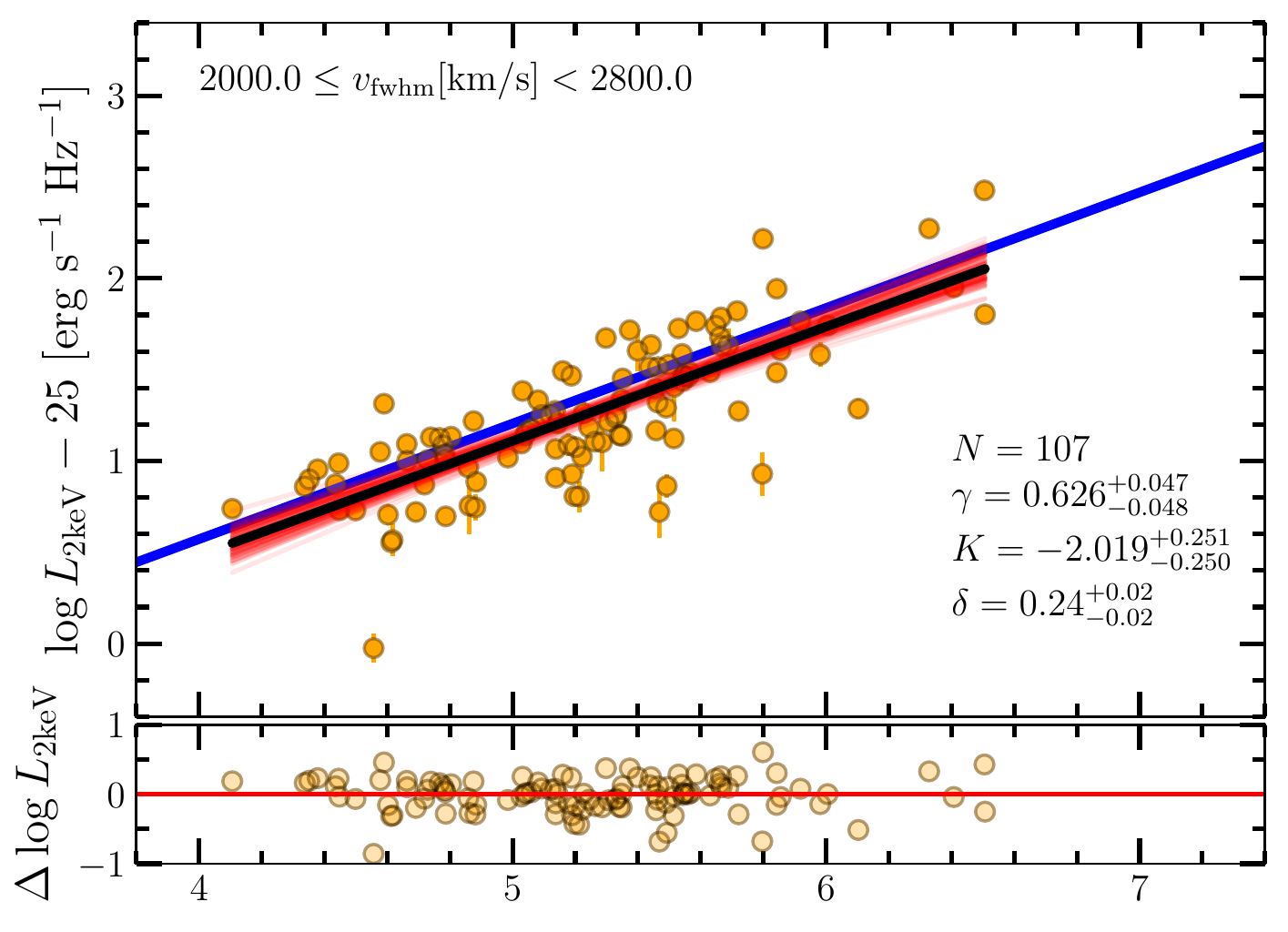}}
 \vspace*{-0.5cm}\subfloat{\includegraphics[width=0.45\linewidth,clip]{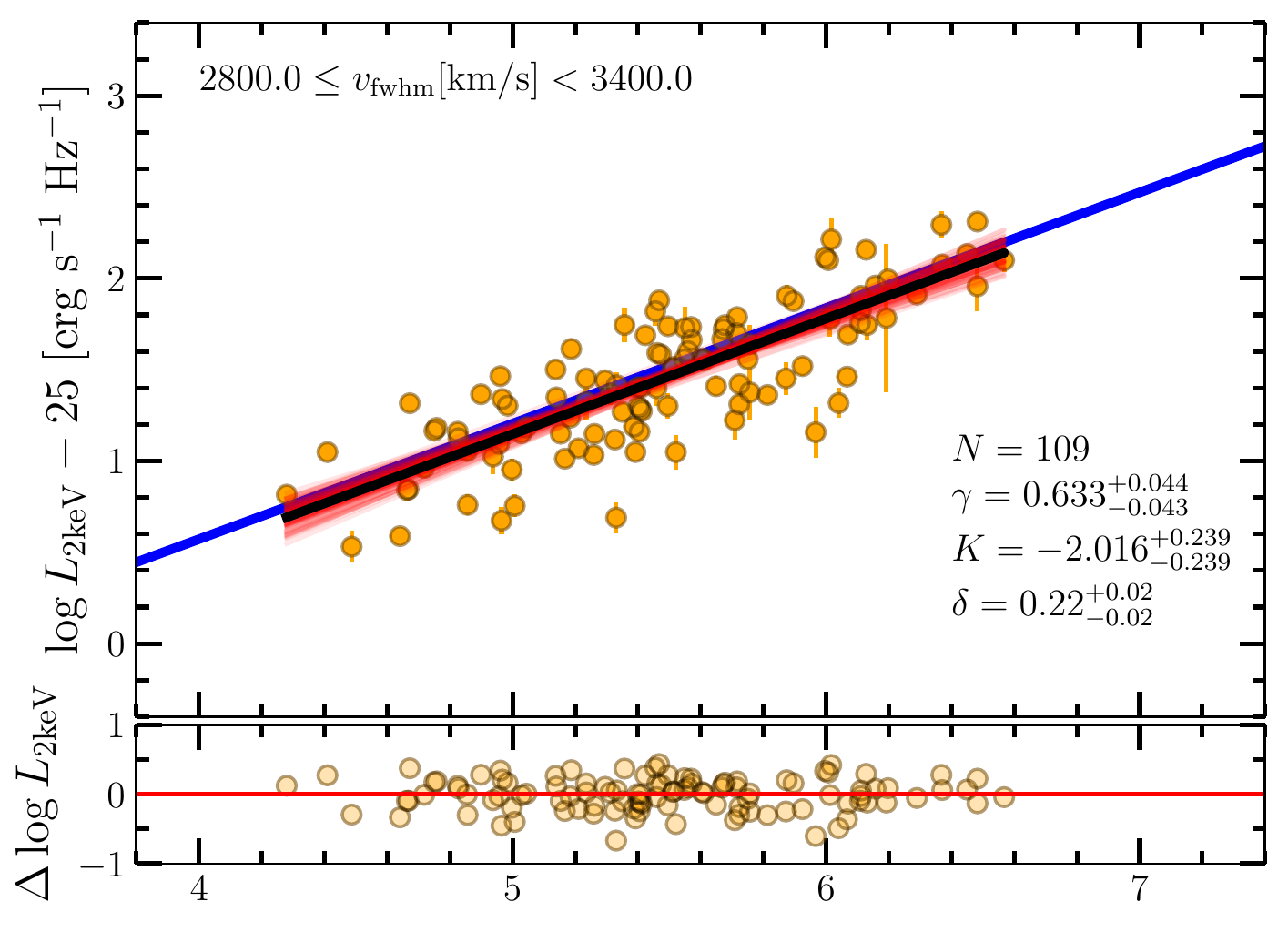}}\\
 \vspace*{-0.5cm}\subfloat{\includegraphics[width=0.45\linewidth,clip]{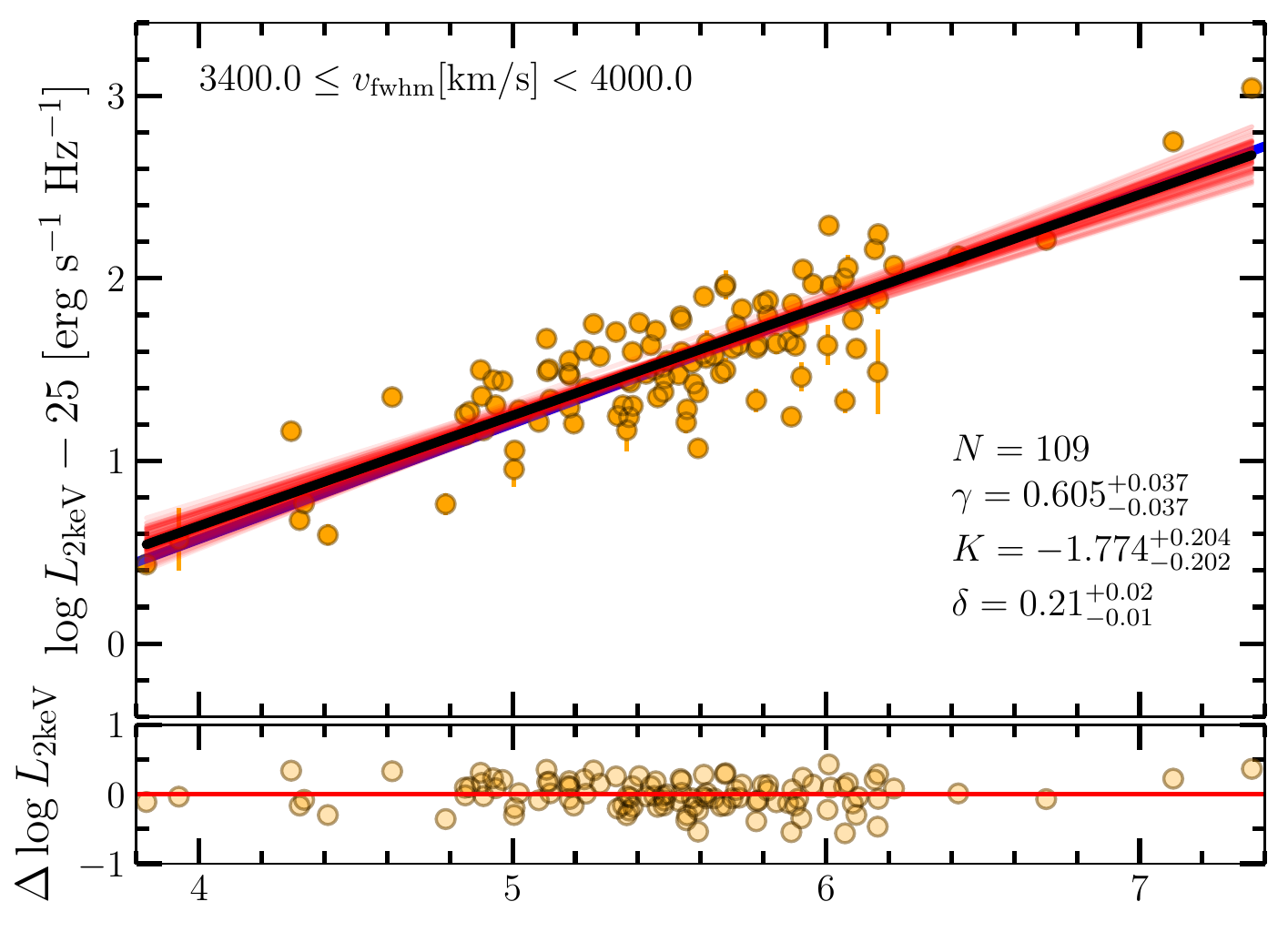}}
 \subfloat{\includegraphics[width=0.45\linewidth,clip]{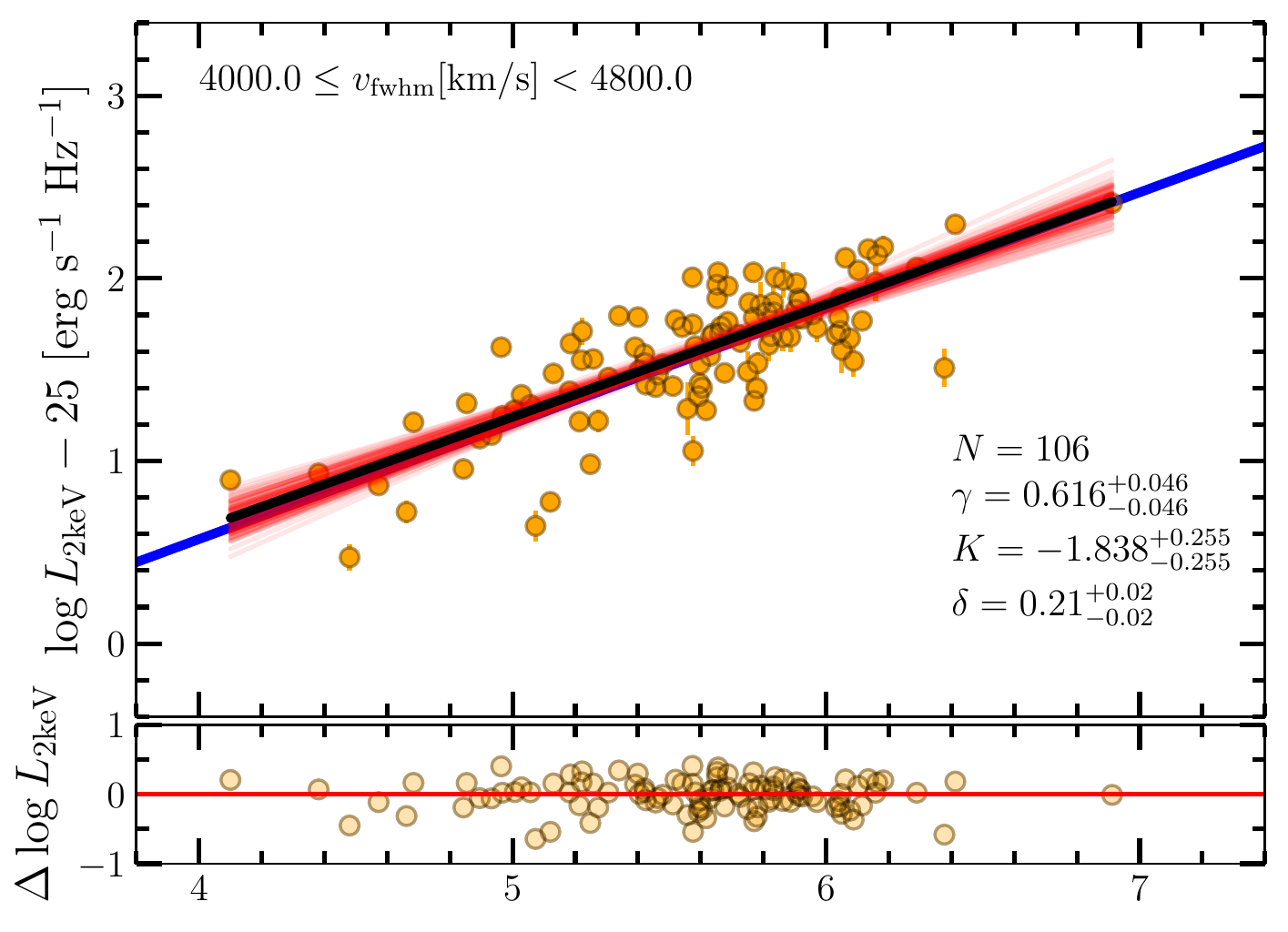}}\\
 \centering\subfloat{\includegraphics[width=0.45\linewidth,clip]{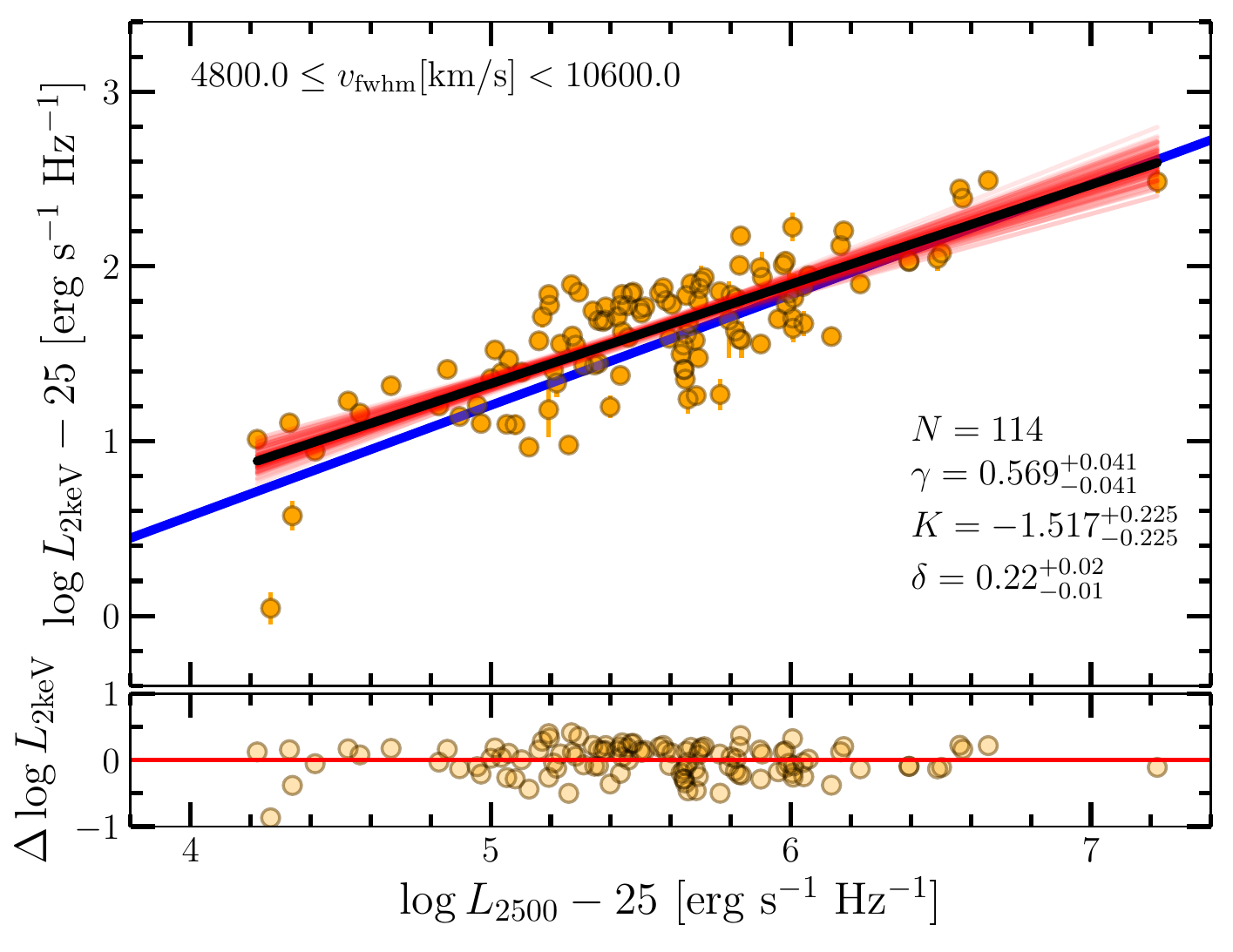}}
 \caption{As in Fig.~\ref{lxloall} in bins (having roughly the same number of objects) of $\w$. The best-fit regression line resulting from the analysis of the whole clean sample (blue solid line) is plotted as a reference. 
 }
 \label{lxlobinw}
\end{figure*}
%-------------------------------------------------------------

\begin{figure*}[t!]
  \subfloat{\includegraphics[width=0.48\linewidth,clip]{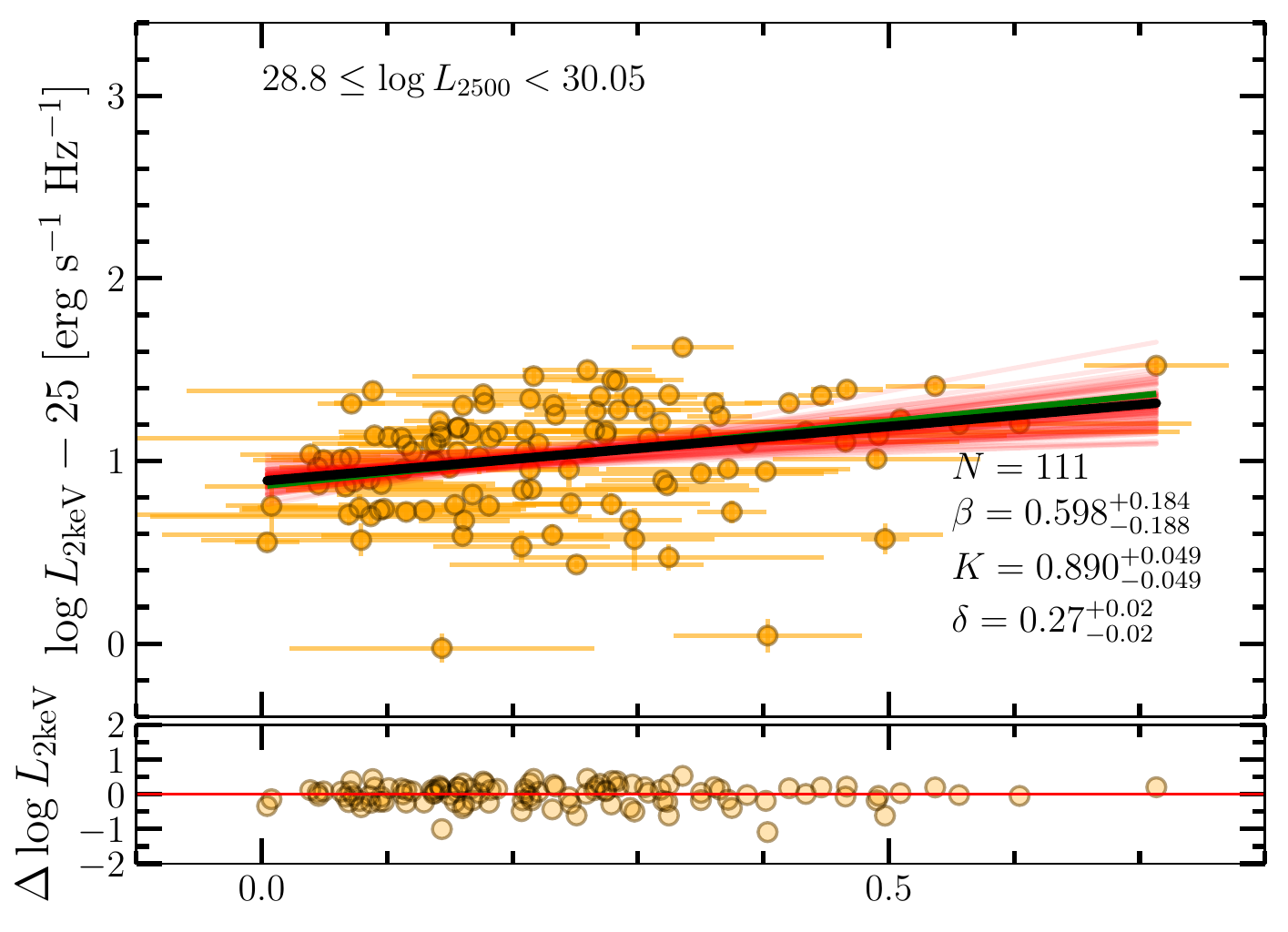}}
  \vspace*{-0.5cm}\subfloat{\includegraphics[width=0.48\linewidth,clip]{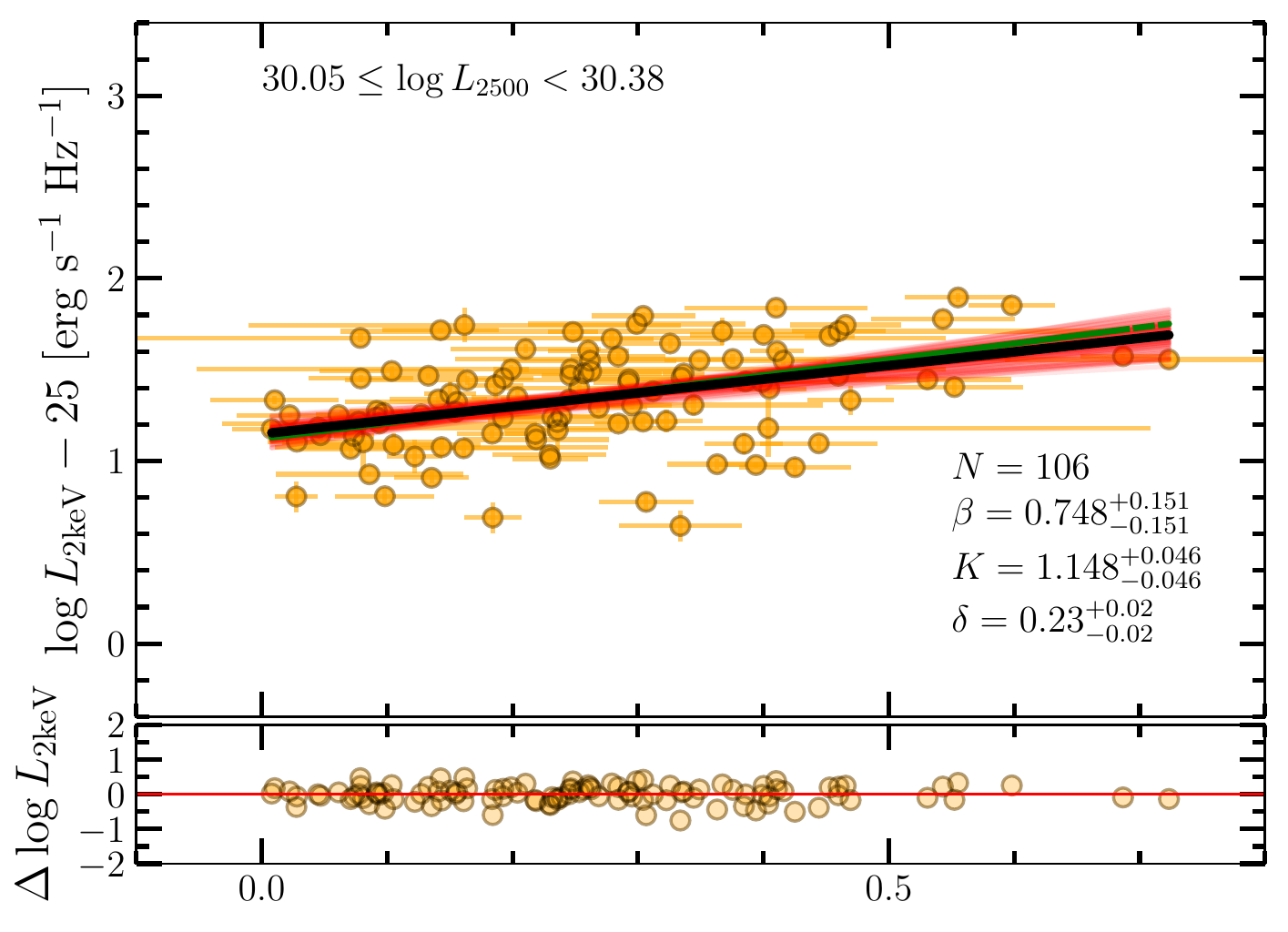}}\\
  \vspace*{-0.5cm}\subfloat{\includegraphics[width=0.48\linewidth,clip]{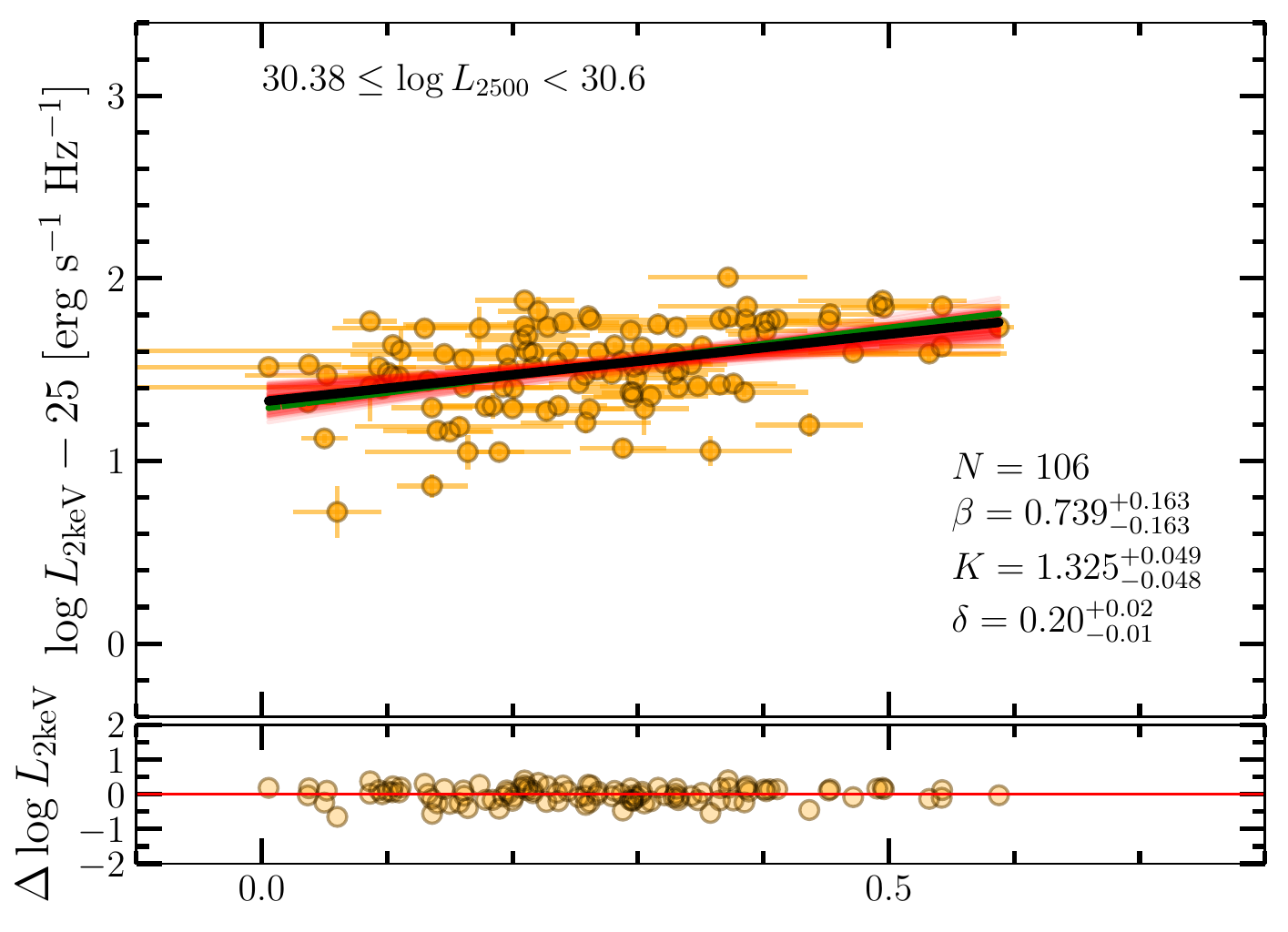}}
  \subfloat{\includegraphics[width=0.48\linewidth,clip]{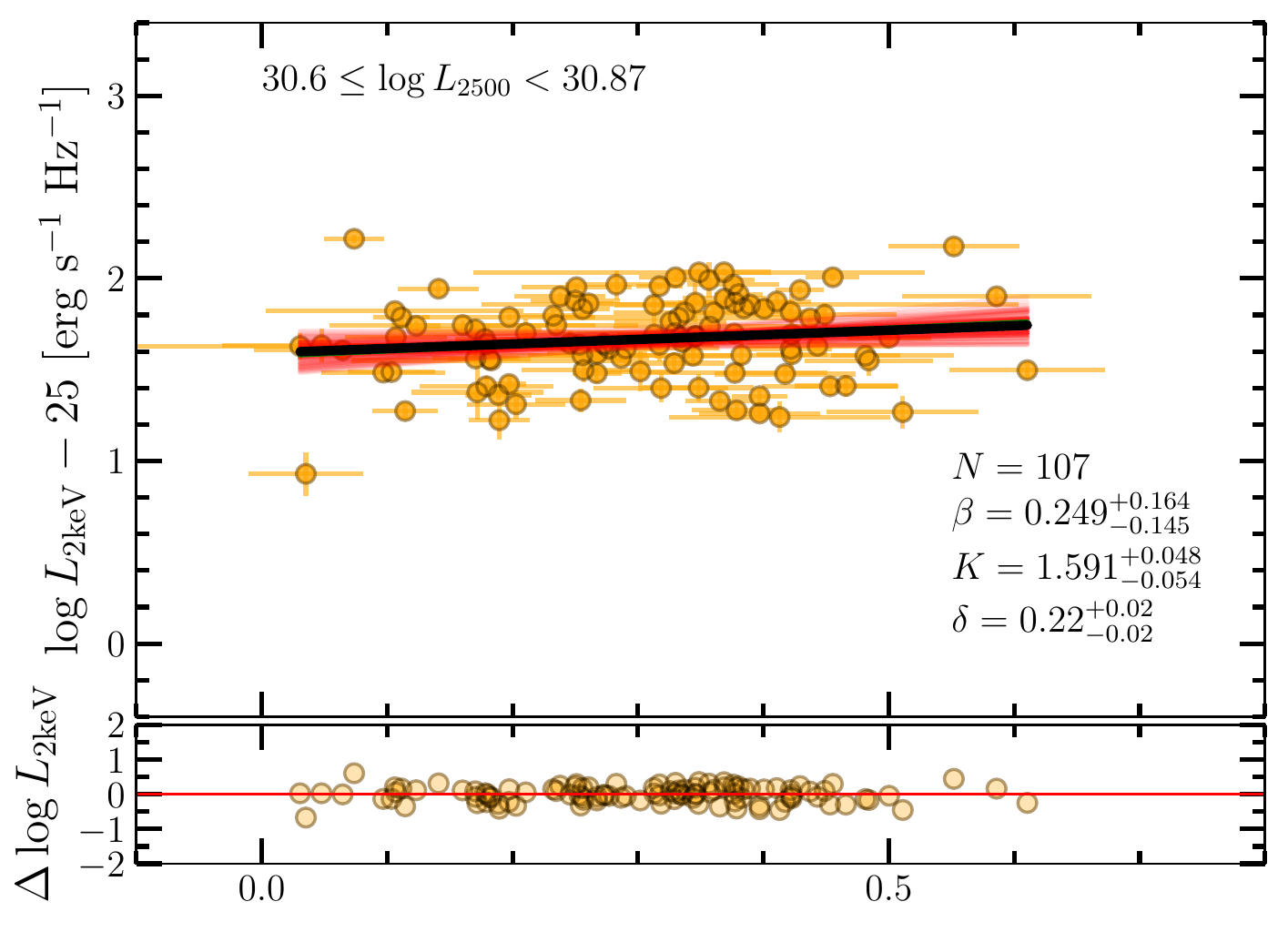}}\\
  \centering\subfloat{\includegraphics[width=0.48\linewidth,clip]{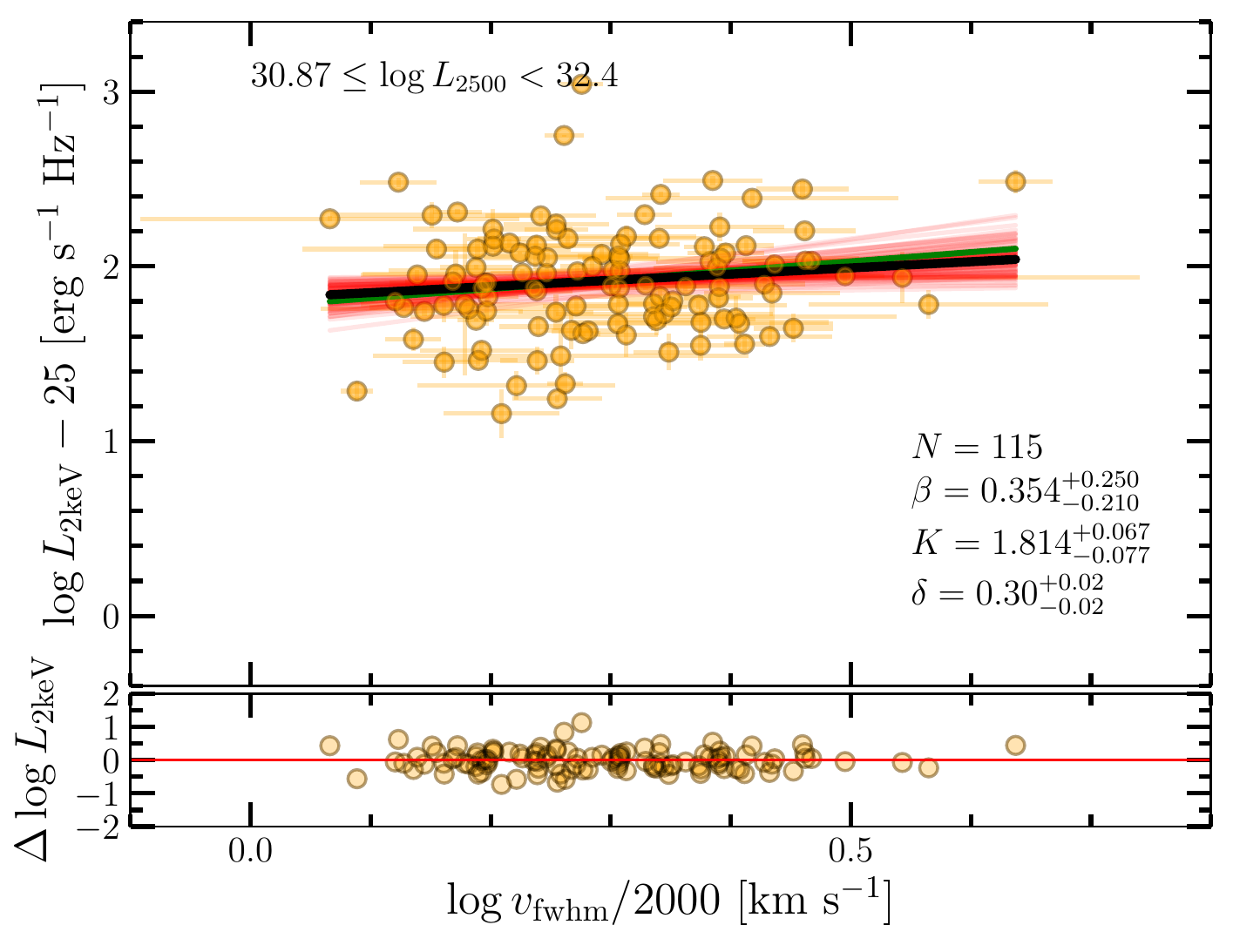}}
 \caption{Rest-frame monochromatic luminosities $\lx$ against $\w$ in bins (having roughly the same number of objects) of $\Lo$.
 Keys as in Fig.~\ref{lxloall}. The best-fit regression line output of LINMIX\_ERR is plotted with the green dashed line for reference.
 }
 \label{lxwbinlo}
\end{figure*}

\begin{figure}
  \includegraphics[width=\linewidth,clip]{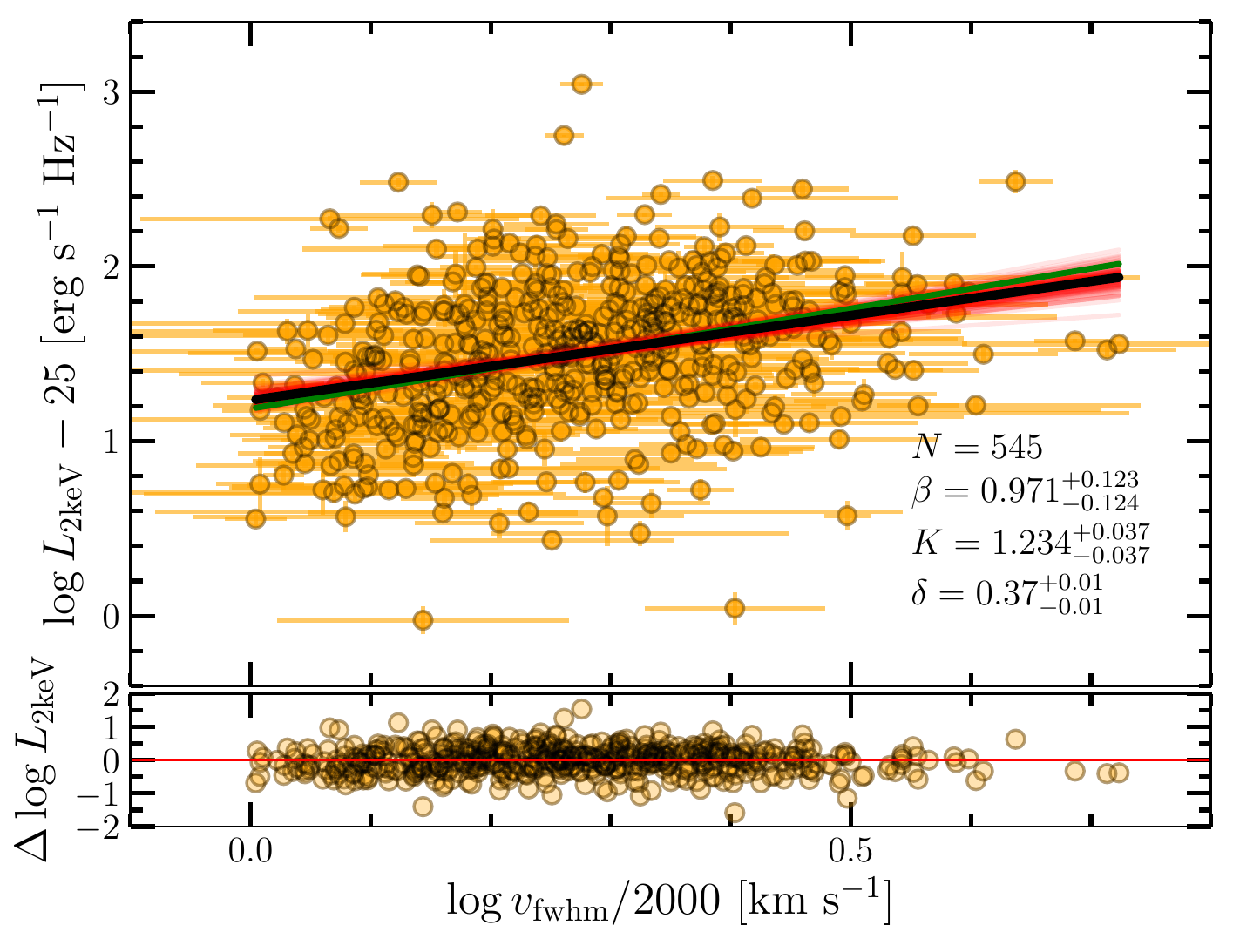}\\
  \includegraphics[width=\linewidth,clip]{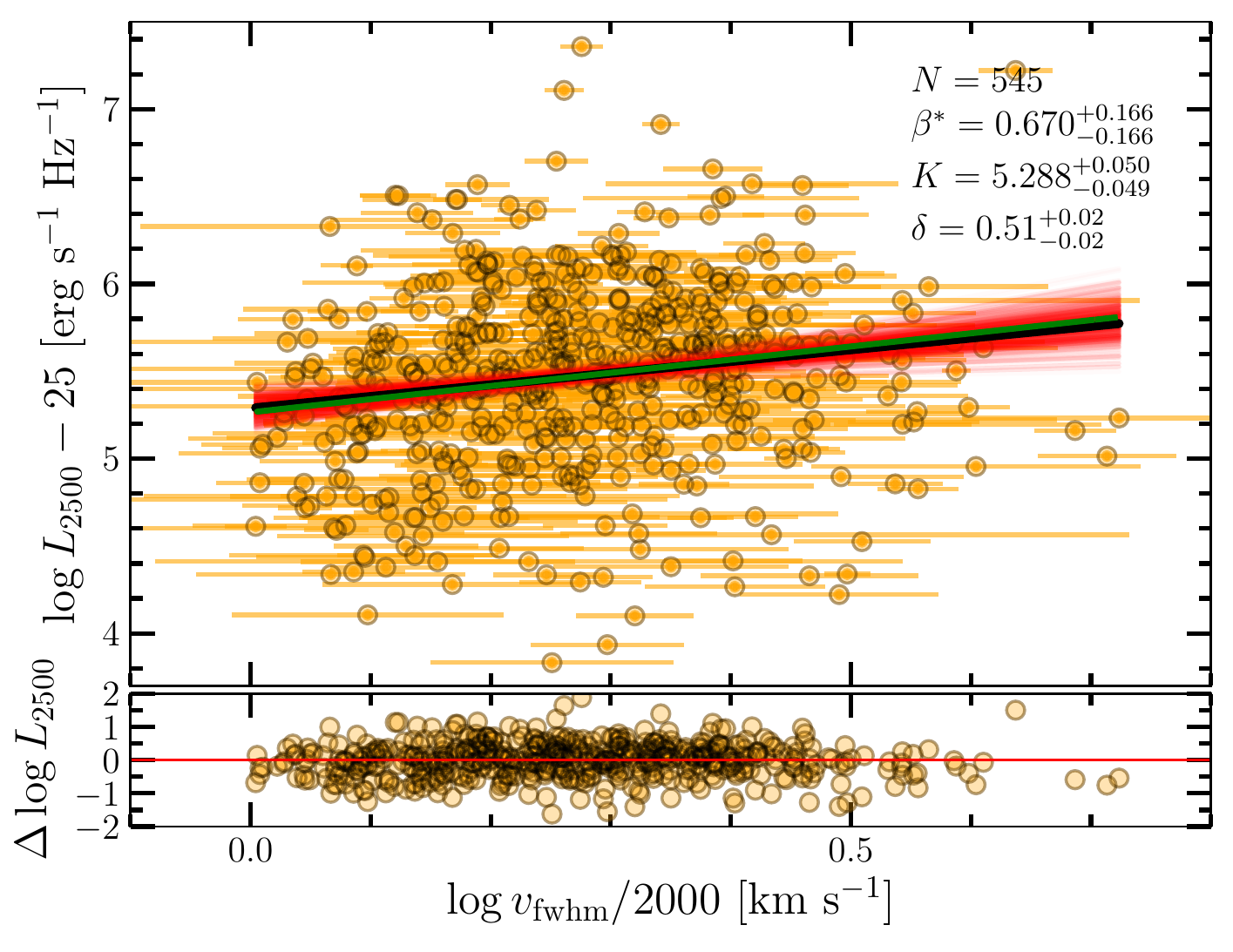}
 \caption{\revs{Rest frame 2 keV and 2500\AA\ monochromatic luminosities as a function of $\w$. Keys as in Fig.~\ref{lxloall}. The best-fit regression line output of LINMIX\_ERR is plotted with the green dashed line for reference. The Student's t-test for the slope of the $\Lo-\w$ and $\Lx-\w$ relations yields a significance that the slope is different from zero at approximately $4$ and $7\sigma$, respectively.}}
 \label{lxw_low}
\end{figure}

\revs{
Figure~\ref{lxlobinw} shows the $\Lx-\Lo$ relationship in five intervals of $\w$ in order to have approximately the same number of sources in each bin.
The slope of the relation is statistically consistent within the uncertainties in all bins with the one of the clean sample as a whole (plotted with the blue solid line for reference). In Figure~\ref{lxwbinlo} we present five cuts of the $\Lx-\w$ relation as a function of $\Lo$. 
Overall, the \revs{estimates} of the slope in each bin present large uncertainties, which are mainly driven by the fact that the $\w$ measurements also display large error bars, thus part of the significance estimated on the observed correlations discussed in Section~\ref{Statistical analysis} may be affected. As a result, we can confirm the correlation between $\Lx$ and $\w$ at the $3-5\sigma$ significance level only in the first three bins, while it is less than $2\sigma$ in the last two intervals. However, our employed fitting techniques perform better than other estimators in cases of large uncertainties (we refer to \citeads{2007ApJ...665.1489K} for a more detailed discussion).

A possible explanation for the tilted residuals in Figure~\ref{lxlowall} is likely due to the combination of high uncertainties on $\w$ and the narrower range covered by $\w$ with respect to both $\Lx$ and $\Lo$. In fact, while $\Lx$ and $\Lo$ cover roughly three orders of magnitude, the majority ($83\%$) of the selected quasars have $\w$ values between 2000 and 5000 km/s in a relatively narrow range of $\Lo$ and $\Lx$. This effect is clear in Figure~\ref{lxw_low} where we show $\Lx$ and $\Lo$ as a function of $\w$ for the selected quasar sample. 
\revs{The Pearson's coefficient for the $\Lx-\w$ and $\Lo-\w$ relations is 0.31 ($t-$value of $\sim7\sigma$ on the slope) and 0.16 ($t-$value of $\sim4\sigma$), respectively. We also point out that the estimated slope of the observed $\Lx-\w$ relation is linked to $\hat\gamma$ and $\hat\beta$ as follows: $\beta=\hat\gamma \beta^\ast +\hat\beta$, where $\beta^\ast$ is the slope of the $\Lo-\w$ correlation.}

\revle{The \revs{median signal-to-noise per resolution element} of the optical spectra in the SDSS--DR7 quasar catalogue within the clean sample is in the range 7--13.}
High signal-to-noise spectra for a larger sample of quasars are required to not only accurately fit the shape of the line, thus improving $\w$ measurements, but also to provide a better leverage at low/high luminosities. 

%\LEt{A and A insists footnotes be incorporated into the body text when discursive and where possible. Please incorporate this footnote into the text }
}

%-------------------------------------------------------------
%-------------------------------------------------------------
%-------------------------------------------------------------
%-------------------------------------------------------------

%%%%%%%%%%%%%%%%%%%%%%%%%%%%%%%%%%%%%%%%%%%%%%%%%%

\end{document}